\documentclass[notrackchanges,twocolumn]{aastex63}
\usepackage{graphicx}
\usepackage{indentfirst}
\usepackage{subfigure}
\usepackage{natbib}
\usepackage{lmacs}
\usepackage{new_aas_macros}
\usepackage{import}
\usepackage[colorinlistoftodos]{todonotes}
\usepackage[capitalize]{cleveref}

\xspaceaddexceptions{]}

\newcommand{\sn}{SN~Ia\xspace}
\newcommand{\sne}{SNe~Ia\xspace}
\newcommand{\deltam}{$\Delta m_{15}(B)$\xspace}
\newcommand{\sbv}{$s_\text{BV}$\xspace}
\newcommand{\vsi}{\ensuremath{v_\text{Si II}}\xspace}
\newcommand{\SiII}{\ion{Si}{2}\xspace}
\newcommand{\Bline}{\SiII~$\lambda5972$\xspace}
\newcommand{\Rline}{\SiII~$\lambda6355$\xspace}
\newcommand{\microns}{$\mu$m\xspace}
\newcommand{\LBol}{$\text{L}_{Bol}$\xspace}

\graphicspath{{./figs/}}

\begin{document}

\title{Extrapolation of Type Ia Supernova Spectra into the Near-Infrared Using PCA}

\author[0000-0002-5380-0816]{Anthony Burrow}
\affiliation{Homer L.~Dodge Department of Physics and Astronomy, University of
             Oklahoma, Rm 100 440 W. Brooks, Norman, OK 73019-2061}

\author[0000-0001-5393-1608]{E.~Baron}
\affiliation{Planetary Science Institute, 1700 East Fort Lowell Road, Suite 106,
             Tucson, AZ 85719-2395 USA}
\affiliation{Hamburger Sternwarte, Gojenbergsweg 112, 21029 Hamburg, Germany}
\affiliation{Homer L.~Dodge Department of Physics and Astronomy, University of
             Oklahoma, Rm 100 440 W. Brooks, Norman, OK 73019-2061}

\author[0000-0003-4625-6629]{Christopher~R.~Burns}
\affiliation{Observatories of the Carnegie Institution for Science, 813 Santa
             Barbara St., Pasadena, CA 91101, USA}

\author[0000-0003-1039-2928]{Eric~Y.~Hsiao}
\affiliation{Department of Physics, Florida State University, Tallahassee, FL
             32306, USA}

\author[0000-0002-3900-1452]{Jing Lu}
\affiliation{Department of Physics \& Astronomy, Michigan State
             University, East Lansing, MI, USA}

\author[0000-0002-5221-7557]{Chris Ashall}
\affiliation{Department of Physics, Virginia Tech, 850 West Campus
             Drive, Blacksburg VA, 24061, USA}

\author[0000-0001-6272-5507]{Peter J.~Brown}
\affiliation{George P. and Cynthia Woods Mitchell Institute for Fundamental
             Physics and Astronomy, Department of Physics and Astronomy, Texas
             A\&M University, College Station, TX 77843, USA}

\author[0000-0002-7566-6080]{James M. DerKacy}
\affiliation{Department of Physics, Virginia Tech, 850 West Campus
             Drive, Blacksburg VA, 24061, USA}

\author[0000-0001-5247-1486]{G.~Folatelli}
\affiliation{Instituto de Astrof\'isica de La Plata (IALP), CONICET,
             Paseo del Bosque S/N, B1900FWA La Plata, Argentina}
\affiliation{Facultad de Ciencias Astron\'omicas y Geof\'isicas
             Universidad Nacional de La Plata, Paseo del Bosque, B1900FWA, La
             Plata, Argentina}
\affiliation{Kavli Institute for the Physics and Mathematics of the Universe
             (WPI), The University of Tokyo, Kashiwa, 277-8583 Chiba, Japan}

\author[0000-0002-1296-6887]{Llu{\'i}s Galbany}
\affiliation{Departamento de F\'isica Te\'orica y del Cosmos,
             Universidad de Granada, E-18071 Granada, Spain}

\author[0000-0002-4338-6586]{P.~Hoeflich}
\affiliation{Department of Physics, Florida State University, Tallahassee, FL
             32306, USA}

\author[0000-0002-6650-694X]{Kevin~Krisciunas}
\affiliation{George P. and Cynthia Woods Mitchell Institute for Fundamental
             Physics and Astronomy, Department of Physics and Astronomy, Texas
             A\&M University, College Station, TX 77843, USA}

\author[0000-0002-4338-6586]{N.~Morrell}
\affiliation{Las Campanas Observatory, Carnegie Observatories, Casilla 601, La
             Serena, Chile}

\author[0000-0003-2734-0796]{M.~M.~Phillips}
\affiliation{Las Campanas Observatory, Carnegie Observatories, Casilla 601, La
             Serena, Chile}

\author[0000-0003-4631-1149]{Benjamin~J.~Shappee}
\affiliation{Institute for Astronomy, University of Hawai’i at Manoa, 2680
             Woodlawn Dr., Hawai’i, HI 96822, USA}

\author[0000-0002-5571-1833]{Maximilian~D.~Stritzinger}
\affiliation{Department of Physics and Astronomy, Aarhus University, Ny
             Munkegade 120, DK-8000 Aarhus C, Denmark.}

\author[0000-0002-8102-181X]{Nicholas~B.~Suntzeff}
\affiliation{George P. and Cynthia Woods Mitchell Institute for Fundamental
             Physics and Astronomy, Department of Physics and Astronomy, Texas
             A\&M University, College Station, TX 77843, USA}

\submitjournal{ApJ}

\received{January 26, 2024}
\revised{March 21, 2024}
\accepted{April 4, 2024}

\correspondingauthor{Anthony Burrow}
\email{anthony.r.burrow-1@ou.edu}

\begin{abstract}

We present a method of extrapolating the spectroscopic behavior of Type Ia
supernovae (\sne) in the near-infrared (NIR) wavelength regime up to
2.30~\microns using optical spectroscopy. Such a process is useful for
accurately estimating K-corrections and other photometric quantities of \sne in
the NIR. Principal component analysis is performed on data consisting of
Carnegie Supernova Project I \& II optical and near-infrared FIRE spectra to
produce models capable of making these extrapolations. This method differs from
previous spectral template methods by not parameterizing models strictly by
photometric light-curve properties of \sne, allowing for more flexibility of
the resulting extrapolated NIR flux. A difference of around
$-3.1$\% to $-2.7$\% in the total integrated NIR flux between these extrapolations
and the observations is seen here for most test cases including Branch
core-normal and shallow-silicon subtypes. However,
larger deviations from the observation are found for other tests, likely due to
the limited high-velocity and broad-line \sne in the
training sample. Maximum-light principal components are shown to allow for
spectroscopic predictions of the color-stretch light-curve parameter, \sbv,
within approximately $\pm 0.1$ units of the
value measured with photometry. We also show these results compare well with
NIR templates,
although in most cases the templates are marginally more fitting to
observations, illustrating a need for more concurrent optical+NIR
spectroscopic observations to truly understand the diversity of \sne in the
NIR.

\end{abstract}

\vspace{30pt}

\section{Introduction}
\label{sec:introduction}

Type Ia supernovae (\sne) are broadly useful tools in cosmology that
provide insight into the nature of dark energy
\citep{Riess_etal_1998,Perlmutter_etal_1999,Brout_etal_2019_a,Brout_etal_2019_b}
as extragalactic distance indicators. This comes as a result of their
standardizable properties; they have been made correctable standard candles in
the past by the use of their light-curve shapes that have been quantified
using empirical parameters such as the decline rate, \deltam
\citep[e.g.][]{philm15,hametal96a}, or stretch-like quantities such as
color-stretch, \sbv \citep{Burns_etal_2014}. Parameterizing these light-curves
facilitates the estimation of the distance modulus and host reddening
through well-known correlations between the intrinsic maximum luminosities of
\sne and their light-curve shapes \citep{philm15,rpk96,philetal99} and colors
\citep{Tripp_1998}.

However, because the precision with which cosmological properties may be
deduced is so dependent on light-curves, the time-evolution of \sne up to
around 40 days past maximum light
must be fully understood to reduce the uncertainty in these cosmological
properties, including Hubble residuals. Unfortunately, the parent systems of
individual \sne remain unclear; neither the progenitor system nor the explosion
mechanism are generally agreed upon within the community
\citep[see, e.g.,][for recent reviews]{Maoz_etal_2014,Jha_etal_2019}. This is
because the carbon-oxygen white dwarfs that explode and produce the spectra of
the \sne we observe are not
observable pre-explosion. As such, through the ever-expanding collection
of recent observations obtained, a common approach to deducing progenitor and
explosion channels is by statistical analysis to seek correlations in both
photometric and spectroscopic indicators in samples of \sne. As more
observations
are obtained, the uncertainties of subtypes of \sne will be reduced,
allowing for better methods of modeling individual \sne, which may improve
their use as cosmological distance indicators and provide more insight into
stellar evolution as a whole.

Recently, the behavior of \sn spectra have been studied by the use of advanced
statistical methods and machine-learning techniques. As sample sizes of observed
\sne are fairly low and time-dependent, it becomes difficult and even
inadvisable to make use of methods such as neural networks in most cases.
However, one approach that handles this limitation relatively well due to its
simplicity is principal component analysis \citep[PCA;][]{Pearson_1901}. PCA is
a process of
dimensionality reduction that is used to describe the highest degrees of
variation in a given sample. This is done by algorithmically calculating a set
of basis eigenvectors --- or principal components (PCs) --- that iteratively
describe the most variation in the sample of data along each subsequent PC
axis.

PCA has been employed to study the diversity of \sne many times in the recent
past. SALT2 and SALT3 \citep{Guy_etal_2007,Kenworthy_etal_2021} use a
technique very similar to principal component decomposition in conjunction by
including a reddening factor to constrain \sn
diversity to two parameters. This allows for estimates of distance and
photometric
redshift as well as aiding in \sn spectroscopic line identification. SNEMO, an
empirical principal-component model has been generated to describe three
different levels of \sn diversity that is comparable to SALT2
\citep{Saunders_etal_2018}. PCA has also been useful in constructing spectral
energy distribution (SED) templates to model the baseline spectroscopic
behavior of various \sne in both the optical and near-infrared (NIR) wavelength
regimes \citep[see, e.g.,][]{Hsiao_etal_2007,Hsiao_2009,Lu_etal_2023}. These templates
have been parameterized by stretch-like quantities such as \deltam
and \sbv, as well as relative phase in time to give them time
dependence. By doing so, these templates provide a convenient inference of an
SED that may be used to determine time-dependent K-corrections to observed
photometry, which are required to correct observed photometry for host-galaxy
redshift \citep{okesand_kcorr68}.

Our primary goal in this project is to employ the use of PCA to create
time-dependent models that describe the variations observed in both optical and
NIR spectra. In this way, it should be possible to provide optical spectroscopy
of any \sn to a model and perform an extrapolation process that infers the
spectroscopic behavior of that \sn at that phase in time in the NIR with some
known
uncertainty. This estimation could be used in a similar manner to templates ---
to estimate an SED in order to perform K-corrections and bolometric luminosity
calculations at any point in time. This is especially useful as optical and
NIR photometry of the same \sn are not often obtained simultaneously,
particularly at early times.

This method is similar to the template method of inferring \sne SEDs
established by \citet{Hsiao_etal_2007} and extended by \citet{Lu_etal_2023}, as
both techniques make use of PCA to formulate a model. However, there are some
key differences that should lead to different results. Most importantly, this
work does not strictly parameterize models by \sbv, but rather by using
abstract correlations in the spectroscopic features themselves. In general,
this may allow more flexibility in how NIR behavior is predicted using optical
information. It may be shown that light-curve shape cannot fully parameterize
spectroscopic or photometric behavior; for example, \citet{Burrow_etal_2020}
show spectroscopically dissimilar \sne that exhibit similar values of \sbv.
\sne with similar \sbv may also show variability of the secondary
photometric light-curve peak seen in the NIR. \citet{Papadogiannakis_etal_2019}
show varying levels of spread in the time of
secondary maximum in the $r$ and $i$ bands for \sne of similar \sbv. Therefore,
it could be that not restricting the parameterization to \sbv may lead to
different and possibly less uncertain results. In this way, this work is
intended to be complementary to \citet{Lu_etal_2023} and provide an alternative
outlook on the problem.

In addition, we also provide an analysis of how our results in the NIR may be
influenced by the classification of a \sn established by \citet{branchcomp206},
which we refer to here as the Branch group classification. These Branch groups
consist of the core-normal (CN), shallow-silicon (SS), broad-line (BL) and cool
(CL) groups, and they have been shown to be robust in a detailed cluster
analysis \citep{Burrow_etal_2020}. This classification system typically uses the
measured pseudo-equivalent widths (pEWs) of \Rline and \Bline to classify
\sne into these four groups with what have been shown in previous studies to be
unique spectroscopic and photometric properties. Photometrically, CNs and BLs
have been shown to exhibit moderate values of \sbv, typically between
$0.7 \lesssim$ \sbv $\lesssim 1.1$, with clearly distinct \Rline features between
them \citep{Burrow_etal_2020}. SSs are usually associated with luminous,
slow-declining \sne, whereas CLs are dimmer and faster-declining. In this paper
we examine how this optical classification scheme may relate to varying
spectroscopic behavior seen in both the optical and the NIR using PCA.

In \autoref{sec:data}, samples and selection criteria used
to inform our PC models are discussed. In \autoref{sec:methods}, the methods
used
to create such models are discussed, as well as how these models are used along
with a given
optical \sn spectrum to predict its NIR spectrum. \autoref{sec:results}
illustrates examples of the prediction process in different cases across
relative time and for different classifications of \sne. Finally in
\autoref{sec:discussion}, we discuss how the resulting PC eigenvectors may
relate to the Branch groups as well as \sbv, and this extrapolation
method is compared with spectral templates, both in spectroscopy and bolometric
luminosity.

\section{Data}
\label{sec:data}

\subsection{Data Source \& Preprocessing}
\label{sec:preprocessing}

To represent the population of observable \sn spectra in both the
optical and NIR, we make use of two different series of spectroscopic \sne
observations. The compiled spectra used in this project are described here.

The optical range of our data is provided by both Carnegie Supernova Project
(CSP) I and II sets of spectra \citep{Folatelli_etal_2010,
Phillips_etal_2019, Hsiao_etal_2019, Burrow_etal_2020}. This combined data set
consists of spectra of 364 unique \sne. The host redshift of \sne in the CSP~I
data set lies in the range of $0.0037 < z < 0.0835$ \citep{Krisciunas_etal_2017},
and that of the CSP~II set is within $0.03 < z < 0.10$
\citep{Phillips_etal_2019}.

The NIR component of the data used here is the same used by
\citet{Lu_etal_2023}, which was obtained using the Folded-port InfraRed
Echellette (FIRE) in the high-throughput prism mode on the 6.5m Magellan Baade
telescope. This instrument captures adequate signal at NIR wavelengths where
considerable telluric absorption typically occurs, allowing for passable
corrections to these regions, albeit still with relatively much noise and
uncertainty. Acquiring as accurate spectroscopic measurements as possible of
\sne in these regions is crucial to achieving an SED suitable enough to
calculate
accurate bolometric light-curves and K-corrections. We choose to use this NIR
set of spectra because this allows for consistency when comparing with NIR
spectral templates of \citet{Lu_etal_2023}, which is a goal of this project. In
addition, the \sne observed with FIRE were purposefully selected to have good
overlap with many CSP II \sne. These FIRE spectra include observations of 142
unique \sne in total; however, using the same selection criteria as
\citet{Lu_etal_2023}, a total number of 94 \sne are included in this NIR
sample. These spectra typically capture up to around 2.4~\microns.

Every \sn used to inform models here also has optical photometric observations
from CSP~I~\&~II \citep{Krisciunas_etal_2017} such that light-curves have
been fit to them using the light-curve-fitting software, SNooPy
\citep{Burns_etal_2014, Burns_etal_2018}. This is so that photometric
quantities required for this analysis, such as time of $B$-maximum and host
reddening, are inferred for each \sn. There are 115 \sne from CSP~I with both
spectra and these photometric quantities available, and similarly 161 \sne from
CSP~II.

Ideally, each spectrum in the data set would be corrected such that its
synthetic photometry would match the observed photometry at the same time. This
may be done using SNooPy, for example, where a best-fit smooth basis spline
with knots at the
effective wavelength of each photometric filter is multiplied by the spectrum
such that the result yields synthetic photometry closest to the observed
values. However, as this project deals with both optical and NIR spectra, this
means simultaneous values of optical and NIR photometry must be available for
each spectrum used. Even with light-curve interpolation for each individual
filter, only a fraction of \sne in our data set have optical photometry
available to allow for this correction, and a much smaller fraction have NIR
($Y$, $J$, and $H$) photometry. Because of the non-parametric nature of the
spline-based color-matching technique, the correction is useful for a study
such as this only when both optical and NIR photometry are available.
Therefore, the spectra used
here are not corrected for observed photometry, and our models do not account
for this source of variation.
However, the CSP~I spectra have been shown to give accurate optical colors
\citep{Stritzinger_etal_2023} including in $r-i$, which includes the wavelength
region where the optical and NIR spectra are merged here.
\citep{Hsiao_etal_2019} also showed that the CSP~II NIR colors obtained from
synthetic photometry were accurate to $0.03-0.08$ mag.

To eliminate as much other known, systematic variation in the training data
sample as possible before the eigenvectors are calculated, some additional
preliminary processing has been performed. Three optical telluric features at
5876--5910~\AA, 6849--6929~\AA, and 7576--7688~\AA\ were removed from each
spectrum. Then, each spectrum was put into the rest frame and corrected for
Milky Way (MW) extinction using reddening estimates of
\citet{Schlafly_Finkbeiner_2011} with $R_V = 3.1$ and a host-galaxy
color-excess value $E(B-V)_{host}$ inferred by SNooPy.
\added{
These corrections were performed using the CCM law given by \citet{CCM_1989}.
}
Note that host-galaxy extinction is less certain than MW extinction, therefore
some variability due to host extinction could be present in PCs described later
in \autoref{sec:methods}.

\subsection{Sample Selection}
\label{sec:sample-selection}

As discussed in \autoref{sec:introduction}, the goal of this extrapolation
method is to use the available data to train a time-dependent model that
describes both optical and NIR spectroscopy. Rather than compiling all
available observed spectra into a single model by the use of some interpolation
technique such as Gaussian processes, we have chosen to generate multiple
models, with each model representing \sn spectra at some point in time.
Specifically, each model will represent an integer day past $B$-band maximum
(i.e. day $-1$, day $+0$, day $+1$, etc.). We chose to incorporate
time-dependence in this way to eliminate possible bias; for example, two \sne
of different subtypes may appear spectroscopically similar in some aspects at
different phases in time \citep[e.g.,][]{Yarbrough_etal_2023}. In doing this,
the effects of time evolution should be minimally captured by our individual
models. In addition, hereafter in this paper, maximum light will always refer
to the time of $B$-band maximum.

The sample of spectra selected from our total data set for each individual
model representing some time $t$ is chosen in the following way. First,
for every \sn that has spectra observed before and after $t$, we interpolate
the spectra using Gaussian process regression (GPR) if there are enough spectra
and there is an observed spectrum close enough to $t$.
\replaced{
We find that if the observed spectra are close enough in time to $t$, that
around 3--4 observations provide a good interpolation with low propagated
uncertainties from the GPR.
}{
We find that 3--4 observed spectra within approximately $-10$ to $+50$ days
of maximum light provide a good interpolation with low propagated uncertainties
from the GPR if at least two of those observation times surround $t$ ---
otherwise it would require extrapolation rather than interpolation.
}
If the \sn is sparsely observed or is otherwise unable to be interpolated, we
choose the spectrum observed closest to time $t$ within $-5$ to $+10$ days
relative to $t$. This window is asymmetric to allow for more spectra to lie
within the window, while assuming observed spectra change more dynamically for
the same \sn at earlier times.
\replaced{
We allow this seemingly large window as splitting up observations into smaller
time windows drastically reduces sample sizes, and we find that the increase in
sample size is worth the small amount of time evolution introduced in favor of
smaller uncertainties, which are discussed in \autoref{sec:uncertainty}.
}{
Although using such a large window may introduce some time evolution into the
model, we find that this is favorable \emph{in lieu} of the smaller sample
sizes that a smaller window would generate. Larger sample sizes would produce
smaller uncertainties, which are discussed in \autoref{sec:uncertainty}.
}

\begin{figure}[ht]
    \centering
    \includegraphics[width=\linewidth]{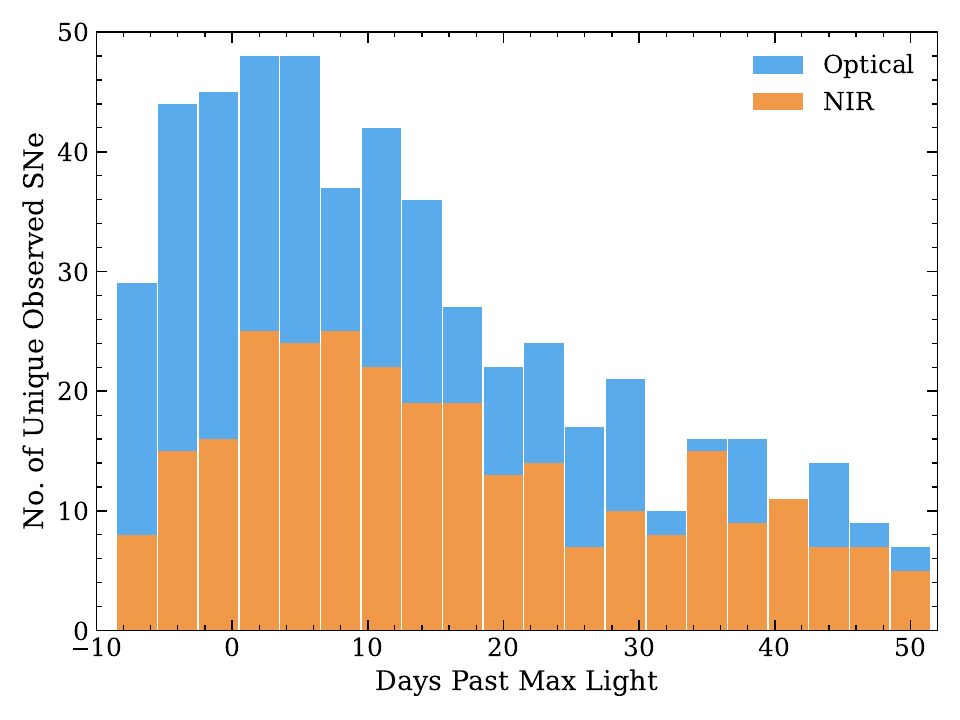}
    \caption{Bars here show the number of unique observed \sne in the optical
             and NIR data sets as a function of time past $B$ maximum in
             three-day bins.}
    \label{fig:sample_distribution}
\end{figure}

\begin{figure}[ht]
    \centering
    \includegraphics[width=\linewidth]{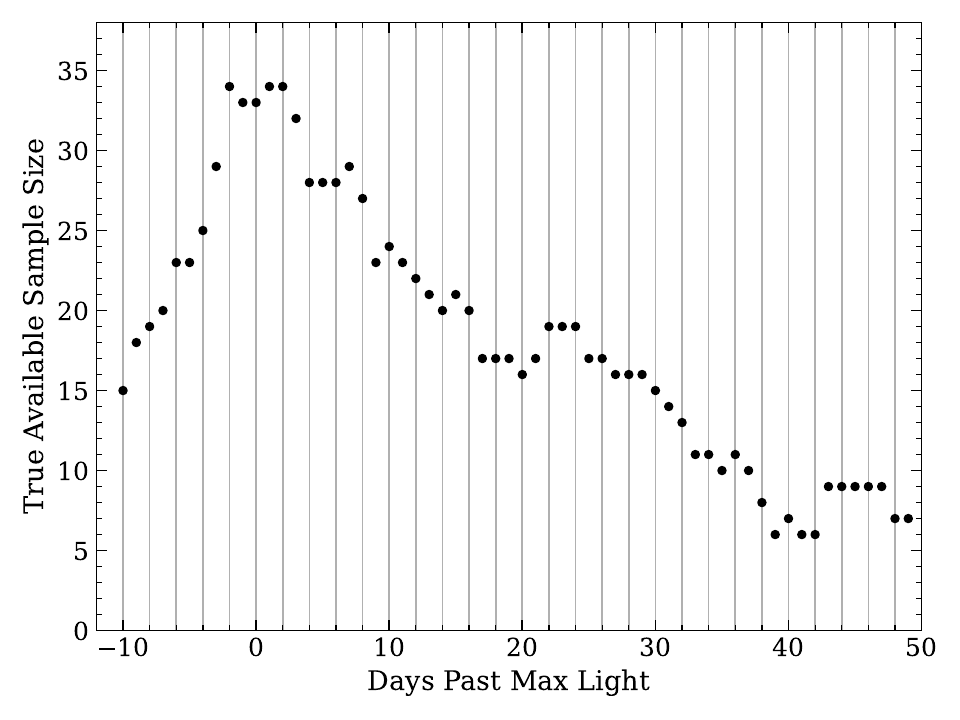}
    \caption{The true available sample size for our models as a function of
             time, accounting for both optical and NIR spectra eligible to have
             a representative spectrum for that time (described in
             \autoref{sec:sample-selection}).}
    \label{fig:sample_times_true}
\end{figure}

To reiterate, small sample size is the primary difficulty in this analysis. To
illustrate the optical and NIR data sets used, in
\autoref{fig:sample_distribution} we show a histogram of the number of unique
\sne from each data set with at least one observation within the given bins,
which are three days wide
\added{
, ranging from $-10$ to $+50$ days relative to maximum light
}.
This only accounts for those we use for the analysis; that is, those that have
had light-curves fit, etc.
\replaced{
The spectra used here typically span around $-10$ to $+50$ days relative to
maximum light, with both samples containing more observations at times near
maximum light. This means that our models will be more applicable for
extrapolating spectra observed near maximum light.
}{
Both samples contain more observations at times near maximum light, implying
that our models will be more accurate when used to extrapolate spectra observed
near the epoch of maximum light.
}
This may be more useful as it is often more difficult to arrange observations
of the same \sne concurrently in the optical and NIR at earlier times, and the
early phase characterizes the light-curves of \sne.

In addition, \autoref{fig:sample_times_true} shows the true sample size
available to each model at each phase $t$ relative to maximum light. This
accounts for the specifications
given in the previous paragraph, including \sne whose spectra may be
interpolated or otherwise are within $-5$ to $+10$ days of $t$. Accordingly,
the sample sizes available for our models at early times are typically around
20--30 \sne, peaking at 34 \sne available when attempting to do an
extrapolation of a spectrum taken around $t = +0$ days. This is still
a low sample size, however PCA is chosen for this analysis because it tends to
work well with low sample sizes due to its simplicity.

Fortunately, the optical and NIR data
\added{
sets
}
have good combined wavelength coverage.
\replaced{
This is ideal, because when a spectrum is chosen from both the optical and NIR
data sets for the same \sn that is representative of some time $t$, the two
spectra must be merged.
}{
This allows for two spectra chosen from the two data sets to be merged into a
continuous spectrum of a \sn representing time $t$ that spans the optical
and NIR.
}
Nearly all optical and NIR spectra have shared wavelength coverage between
0.81--0.84~\microns. Choosing the higher value
\replaced{
here
}{
of this range
}
to remove optical
fringing in the NIR data, we choose to scale the NIR spectrum by a constant
factor that allows the two spectra to be equal at 0.84~\microns. As the
optical and NIR spectra are separately smoothed prior to merging, described in
more detail in \autoref{sec:methods-pca}, noise around 0.84~\microns should not
affect this scaling factor much. The two spectra are then merged into a full
optical+NIR spectrum for that \sn, ignoring everything redward of 0.84~\microns
for the optical spectrum and blueward
of 0.84~\microns for the NIR spectrum. This entire process is performed for
each supernova until a complete sample of spectra representative of each
time $t$ that span both the optical and NIR is achieved.

\added{
As stated in \autoref{sec:preprocessing}, the individual optical and NIR
spectra are unable to be consistently color-corrected. Consequently, each final
merged spectrum is also not corrected to match the photometric colors. We
discuss potential impacts of this on the results in more detail in
\autoref{sec:color-matching}.
}

\section{Methods}
\label{sec:methods}

\subsection{PCA Method}
\label{sec:methods-pca}

PCA is a dimensionality-reduction technique that allows one to describe the
variation in a sample of training data made up of $N$ vectors of the same
dimension. In this case, the sample here is a set of compiled spectra described
in \autoref{sec:sample-selection}. To transform each spectrum into a vector in
the same space, each spectrum in the training sample is interpolated to the
same points in wavelength space. This interpolation is performed via a Gaussian
process regression using Spextractor \citep[see][for a more in-depth discussion
of this process]{Burrow_etal_2020}. As a result, the spectra are also smoothed,
eliminating much of the noise each spectrum may contain.

To account for any spectra in the sample that have relatively high flux
uncertainty, for example in telluric regions, a variant of standard PCA called
expectation-maximization PCA (EMPCA) \citep{Roweis_1997,Bailey_2012} is used.
EMPCA uses estimates of uncertainty of input data to weight each sample such
that the variation described by eigenvectors is not dominated by high-noise
samples or extreme outliers. It is also particularly useful when working with
data sets with missing points.

Typically one weights each point by a quantity similar to the inverse variance
$1 / \sigma^2$; however, in our testing we found this particular weighting
scheme leads to noisy eigenvectors, possibly due to the fluctuation in the
errors from point to point on the spectra. This does not significantly affect
the end results when reconstructing spectra with more eigenvectors (discussed
in \autoref{sec:methods-prediction}), however when attempting to reconstruct
spectra with fewer eigenvectors, the predictions become much noisier. We
instead make use of a weighting scheme that is linear in variance, such that
the weight of the $i$th wavelength point of each sample is given by
$1 - 0.75\ \sigma^2_i / \sigma^2_{i, max}$, where $\sigma^2_{i, max}$ is the
largest variance exhibited at that wavelength across all samples. The factor of
0.75 is inserted to prevent high-variance data from being given zero weight,
and it is an arbitrary parameter that describes the relative weighting between
high- and low-variance data points. This min-max scaling of flux variance
chosen for the weighting scheme is not standard, but we find it works well in
limiting the weight of spectra containing relatively high uncertainties.

\subsection{PC Training}
\label{sec:training}

Once the sample spectra are interpolated in wavelength space
\added{
(as discussed in \autoref{sec:methods-pca})
},
applying PCA will
effectively be looking at how each individual wavelength point varies for each
spectrum in the sample, and connecting them to form spectrum-like eigenvectors.
Because of this, we may choose any set of wavelength points for our training
data. However, to be consistent and eliminate as many variables from this
analysis as possible, we choose to use spectra only in the range of
0.50-2.30~\microns. We typically do not use data blueward of this region
because we find in testing that including Fe-group features blueward of
0.50~\microns adds too much variation to the training sample given the
available sample sizes
\added{
, leading to lower quality fits and less consistent predicted results
}.
This variation stems from sources such
as line
blanketing and Ni mixing
\citep[see, e.g.,][]{Baron_etal_2006,Hoeflich_2017,DerKacy_etal_2020}.
\added{
The exact reason that adding more information to the model blueward of
0.50~\microns leads to inferior results is unknown. However, this seems to
suggest that the features in this bluer optical region are not correlated with
much of the NIR in a way that PCA can describe easily with the limited data
available. Perhaps studying the variation in the UV+optical region is the
subject of future work when more data is available, but with our current
data limitations it is outside the scope of this project.
}

Once a set of spectra that represent the same time and the same points in
wavelength space is compiled, they need to be normalized and standardized
before performing PCA. First, each spectrum is individually normalized by its
mean value of flux between 0.50--0.84~\microns. This is chosen because
normalizing by the mean (rather than, for example, the maximum flux value)
eliminates some of the variation between spectra due to intrinsic differences
in color or other related quantities. It is important to note that PCA is quite
dependent on the choice of normalization used, so another normalization may
produce
\deleted{
significantly
}
different results.

After this normalization, the training data are all uniformly standardized.
This includes subtracting the total mean spectrum $\mu_F$ of all spectra in the
sample, and subsequently dividing them by the standard deviation of the sample,
$\sigma_F$, such that the variation of flux in the sample at each wavelength
point has a standard deviation of 1. This ensures that the eigenvectors are
describing true variation from the mean, and that each wavelength point is on
the same scale. This is important because there would be a larger magnitude of
variation in the optical part of a spectrum than the NIR region, as spectra
typically exhibit higher values of optical flux.

\subsection{Prediction Using PCA Eigenvectors}
\label{sec:methods-prediction}

EMPCA produces a basis of eigenvectors that describe most of the variation in
the training data sample.
\deleted{
We assume that this basis may be used to describe the optical-to-NIR spectra of
the total population of \sne.
}
\replaced{
It is also assumed that there will be either physical or non-physical
correlations between different regions in wavelength space that the
eigenvectors will describe.
}{
It is desirable for these eigenvectors to capture either physical or
non-physical correlations between different regions in wavelength space.
}
As an example, the third PC may describe that the sample typically exhibits
increased O~I and NIR Ca~II simultaneously, which would hint that the two
features may be correlated to some extent for \sne. Indeed this is the case for
spectra at maximum light, as shown in \autoref{sec:results-eigenvectors}. A
more in-depth discussion of the behavior of eigenvectors for maximum-light
spectra is also given in \autoref{sec:results-eigenvectors}.

By making use of these correlations or anti-correlations, if the spectroscopic
behavior of a \sne can be predicted in the NIR using an observed optical
spectrum, the optical spectrum can be projected onto the optical part of each
eigenvector in the basis to get a set of eigenvalues. This projection is done
via a simple dot product, treating the spectrum and eigenvectors as vectors
in wavelength space. It should be noted that the spectrum first needs to be
normalized and standardized in the exact same manner as the training data
(discussed in \autoref{sec:methods-pca}). The resulting projection values just
correspond to the eigenvalues needed to reconstruct the optical part of the
spectrum --- which is already known --- using the eigenvector basis. In other
words, these projection values describe the extent that each eigenvector
describes the optical spectrum. To extrapolate the spectrum in the NIR
region, the optical eigenvalues may then be used with the full eigenvectors
that span both the optical and NIR regions to form a linear combination that
reconstructs the spectrum in both the optical and NIR.
\replaced{
This process assumes that there are eigenvectors that describe variation in
both the optical and NIR simultaneously. Because the data have been
standardized, this should generally be the case, as the eigenvectors describe
unit-scaled variation of the spectra at each wavelength.
}{
This is possible because each eigenvector individually describes both optical
and NIR variation simultaneously. Because the data have been standardized,
the variation seen in the optical and NIR will be on the same unit-scale.
}
This process may also be performed with any subset of the eigenvector basis,
but we assume that the reconstruction is more precise with more PCs used, as
each PC accounts for
\added{
further
}
variation in the sample.
\replaced{
This is assuming the PC is not capturing pure noise, as could be the case for
some higher-degree PCs.
}{
However, higher-degree PCs could be capturing pure noise and are therefore
often irrelevant to much of the analysis and the prediction.
}

\subsection{Uncertainty Measurement}
\label{sec:uncertainty}

A measurement of uncertainty for these predictions is given by accounting for
two different sources of variation. The first of which comes from uncertainty
in how well the eigenvectors are able to be used to make the exact same
predictions for the training sample itself, including the same region projected
onto the eigenvectors to make a prediction. More specifically, the same
prediction is made for each \sn in the training sample, and then the sample
standard deviation of these predictions is used as part of the total
uncertainty. This treats each wavelength point as purely uncorrelated, that is,
the noise is treated as white noise. However, the wavelength points in the
spectrum are correlated by the underlying physics. In a future release of our
code, we will correct for this correlation, but it is beyond the scope of the
present work.

Secondly, if there is uncertainty in the flux in the known part of the spectrum
that is being extrapolated, this could affect the prediction and therefore must
be represented in the total uncertainty. To account for this, we assume that
the flux uncertainty is distributed normally and draw 500 samples of the
spectrum. Using these samples we then make the same predictions using the same
projection and prediction regions, and the standard deviation in these
predictions is calculated. This is then added in quadrature to the previous
source of uncertainty to make up the total uncertainty in the reconstructed
spectrum.

\section{Results}
\label{sec:results}

\subsection{Maximum-Light Eigenvectors}
\label{sec:results-eigenvectors}

\begin{figure*}[ht]
    \centering
    \includegraphics[width=\textwidth]{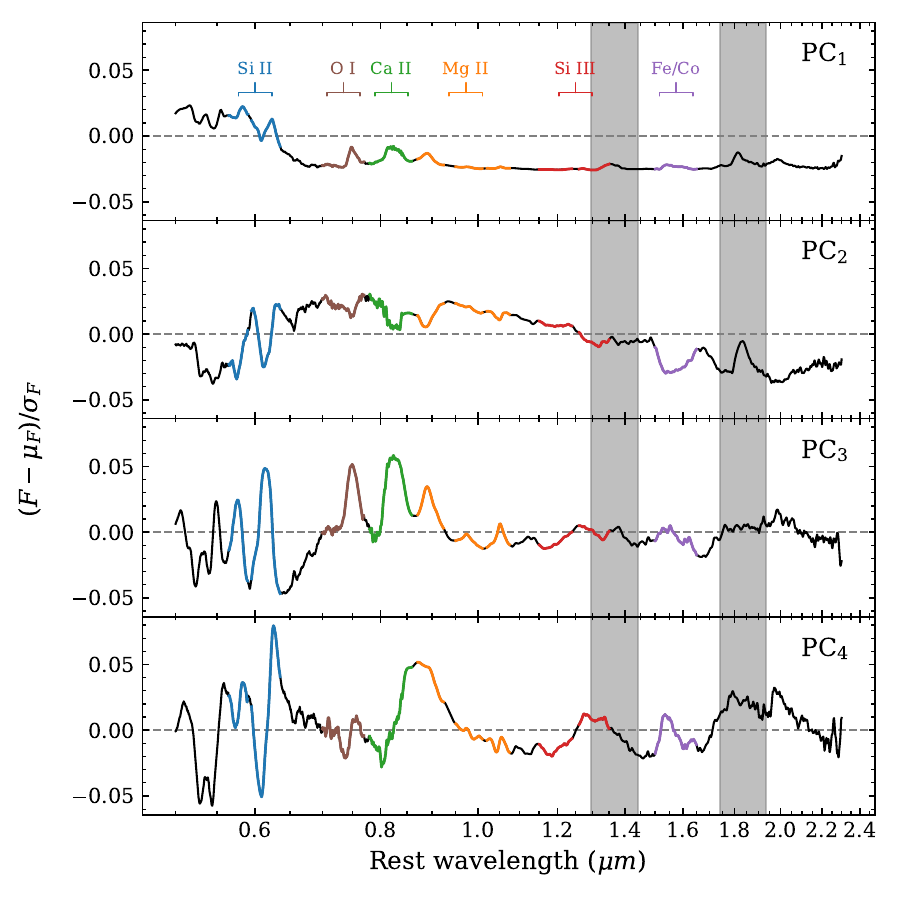}
    \caption{The first four PC eigenvectors representing $t=0$ that describe
             the highest degrees of variation in the near-maximum-light
             standardized training data set. These vectors lie within
             wavelength space, and note that these eigenvectors represent
             unit-scaled variation from the mean spectrum. Grey shaded
             regions indicate the larger telluric regions in the NIR.}
    \label{fig:eigenvectors}
\end{figure*}

When a standardized training data set has been established to represent some
point in time, a model may be created that includes a set of N eigenvectors
which describe the N highest degrees of variation around the mean of that
sample of spectra. \autoref{fig:eigenvectors} shows the first four
eigenvector spectra that describe the highest amount of variance in the
training sample of 33 unique \sne spectra observed at or around maximum light
($t=0$) given by the EMPCA algorithm. Note again that these eigenvectors are
standardized, showing unit-scaled variation around the mean spectrum $\mu_F$.

In \autoref{fig:eigenvectors}, it can be seen that the first PC seems to
represent a broad correction to spectra. This could be related to some
stretch- or color-like quantity; however, this is unclear as later in
\autoref{sec:sbv}  it can be seen that the first PC is not exactly correlated
with the color-stretch quantity \sbv, but rather a linear combination of PCs
may describe \sbv. The other PCs seem to describe more individual
feature behavior. The typical location of some of the major \sn spectroscopic
features are indicated and colored in \autoref{fig:eigenvectors}. Because of the
abstract nature of these eigenvectors, it is unclear visually if an individual
PC captures all information about individual features. For example, both the
third and fourth PCs seem to explain variation around \Rline, although one may
be explaining its depth or pseudo-equivalent width while the other makes
corrections in accordance with
the \Rline velocity extent. This is discussed in a more quantitative manner in
\autoref{sec:pcs_branch}.

\begin{figure}[ht]
    \centering
    \includegraphics[width=\linewidth]{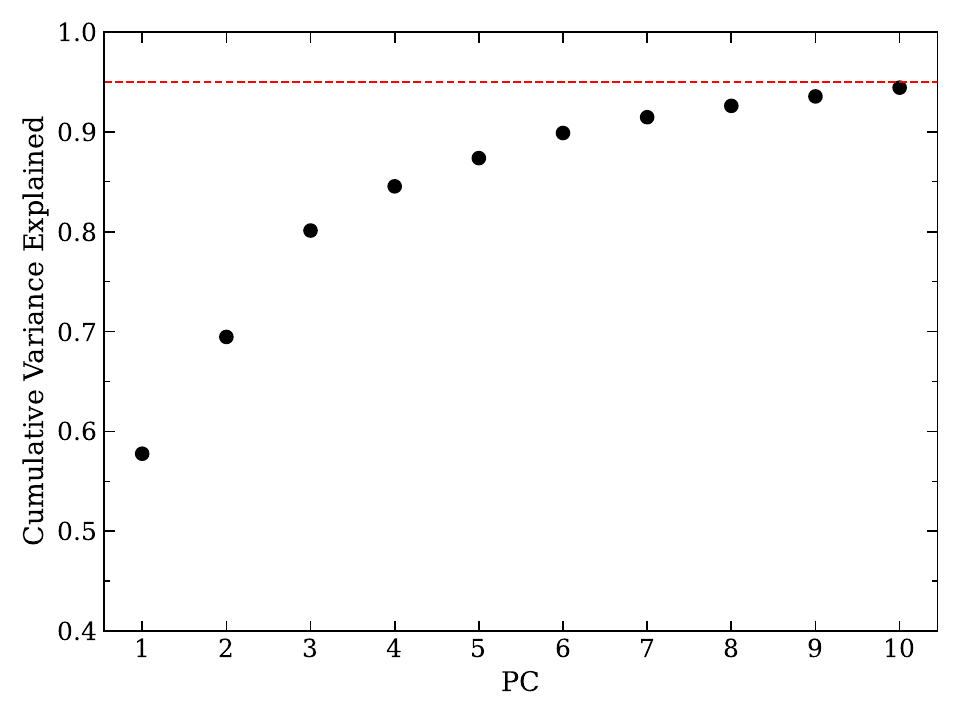}
    \caption{Cumulative fractional variance of the training spectra explained
             by the eigenvectors in \autoref{fig:eigenvectors}. The red dashed
             line represents when a total of 95\% explained variance is
             achieved.}
    \label{fig:explained_variance}
\end{figure}

The cumulative fraction of explained variance for each subsequent eigenvector
is shown in \autoref{fig:explained_variance}. As may be seen, 10 eigenvectors
explain up to 94.8\% total variation of the training sample.
\replaced{
We assume that because the spectra have all been standardized --- set on the
same scale at each wavelength --- that 10 eigenvectors is sufficient in
reconstructing typical spectra belonging to the same population of \sne.
}{
Because the spectra have been standardized at all wavelengths, this
demonstrates that 10 eigenvectors is sufficient in reconstructing typical
spectra belonging to the same population of \sne.
}

We speculate that much of the variance that is unexplained by the PCs may stem
from the relatively high uncertainty in the corrections to NIR telluric regions
made for the FIRE data set. In the near future, the negative impact of telluric
features on data driven models may be mitigated with JWST observations
of \sne
\citep[see, e.g.,][]{Kwok_etal_2023}.
% \authorcomment1{Replaced Derkacy et al., etc. with Kwok et al.}
%
% \replaced{
% \citep[see, e.g,][C. Ashall et al., in prep.]{DerKacy_etal_2023aefx,DerKacy_etal_2024xkq}.
% }{
% \citep[see, e.g.,][]{Kwok_etal_2023}.
% }

\subsection{Prediction Near Maximum Light}
\label{sec:results-max-light}

\begin{figure*}[ht]
    \centering
    \includegraphics[width=\textwidth]{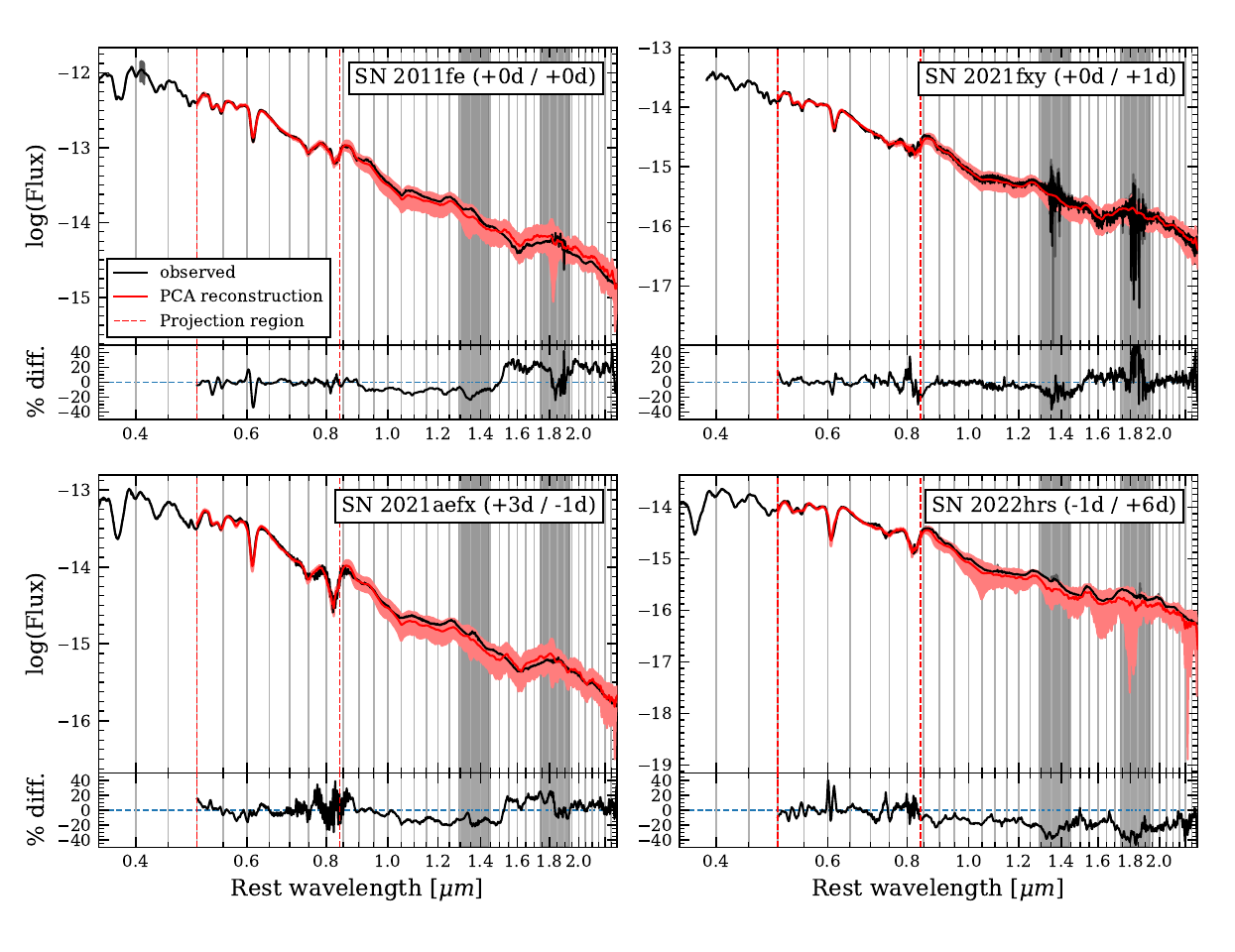}
    \caption{The extrapolation of different \sne near maximum light shown in
             red, compared with the true observed spectrum in black.
             Uncertainties in the predictions are given by the red shaded areas.
             Grey shaded areas specify telluric regions in the NIR. The region
             between red dashed lines is projected onto the maximum-light
             eigenvectors to make this prediction. Percent-difference between
             the prediction and flux is given on the bottom of each panel;
             note this difference is not calculated from log-space. Phases
             shown are the phases of the optical and NIR spectra, respectively,
             that were merged to create the observed spectrum.}
    \label{fig:single_combined}
\end{figure*}

Unfortunately, there is much complexity in the time- and wavelength-dependence
of these predictions. There are also not many homogeneously observed spectra
available in both the optical and NIR at nearby phases in time outside the
sample used in the PCA here. Because of this, a formal test and quantitative
statistical analysis of how well this method actually performs is difficult to
achieve. Instead, to illustrate the potential of this method, in this section
we show a few near-maximum-light examples of this extrapolation procedure. In
particular, we show this in \autoref{fig:single_combined} for SNe 2011fe,
2021fxy, 2021aefx, and 2022hrs, which are a few diverse,
well-observed \sne that are all outside the training sample used to create
the maximum-light eigenvectors shown in
\added{
\autoref{fig:eigenvectors} and discussed in
}
\autoref{sec:results-eigenvectors}.
Because the full optical--NIR range of these \sne is required here for
comparisons, the closest optical and NIR observations to maximum light
available to us were merged into the black line shown in
\autoref{fig:single_combined}. It should be noted that some of these NIR
spectra were observed days apart from their optical counterparts, which may
affect results, though this is not avoidable due to data limitations.

Each of the panels of \autoref{fig:single_combined} show a prediction for
0.50--2.30~\microns indicated in red. They are compared with their respective
rest-frame observed spectrum in black, part of which was used to generate a
prediction. These observations are referenced and further elaborated on below.
Vertical dashed lines indicate the 0.50--0.84~\microns region that is projected
onto the PC eigenvectors to make this prediction. A measurement of uncertainty
for the prediction (see \autoref{sec:uncertainty}) is shown in the light red
shaded area. The two NIR telluric regions between 1.2963--1.4419~\microns and
1.7421--1.9322~\microns are shaded in gray. Percent-differences between the PCA
reconstructions and observed spectra are shown below each spectrum; note that
these differences are in flux space --- not in log-flux space.

The top-left panel of \autoref{fig:single_combined} shows the extrapolation of
a near-maximum spectrum of SN~2011fe \citep{Mazzali_etal_2014_11fe}.
Calculating \SiII line velocity \vsi, pEW(\Rline), and pEW(\Bline) using this
spectrum, and using
values with Gaussian mixture models (GMMs) from \citet{Burrow_etal_2020},
SN~2011fe is classified to be a CN \sn with a probability of 74.2\%. Because
SN~2011fe has historically been so well-observed and studied, this \sn is
included here to illustrate our extrapolation performance on a quintessential
test case using this spectroscopically normal \sn.

% 11fe
% Core-normal, 74.15%
% Median past 1.5u : 18.789
% Median in 0.9u < lambda < 1.5u : -10.453
% Integral F * lam % difference : -3.083
% Si II
% Median past 1.5u : 19.586
% Median in 0.9u < lambda < 1.5u : -9.604
% Integral F * lam % difference : -2.366

\replaced{
For SN~2011fe, the prediction fits the observed spectrum quite well. Within the
region projected onto the eigenvectors (indicated by dashed lines), the PCA
reconstruction of the spectrum performs well, as expected, and the uncertainty
is relatively small.
}{
Looking at the prediction for SN~2011fe in \autoref{fig:single_combined}, the
reconstruction of the spectrum performs well within the region projected onto
the eigenvectors (indicated by dashed lines), and the uncertainty is relatively
small.
}
There is at most a nearly 35\% difference in the flux of
observation and prediction in this region where the depth of \Rline is not
fully reproduced. Outside this region --- past 0.84~\microns where the
extrapolation actually occurs --- the general trend of flux is captured by the
prediction, as well as most individual feature behaviors outside the two large
telluric regions in the NIR. However, in this region the uncertainties are much
larger. In addition, at around 1.5~\microns the prediction begins to deviate
from the observations. Between 0.9--1.5~\microns, the median percent-difference
in the flux is $-10.5$\% --- not including flux in telluric regions, which will
continue to be ignored for other median values provided. After 1.5~\microns the
same median jumps to $+18.8$\% difference in the flux.
\replaced{
Although these figures seem high, when integrating the quantity
$\lambda F_\lambda $ across the entire spectrum outside of telluric regions,
which gives perhaps a better merit of quality with use for corrections to
photometry, the difference is only $-3.1$\%.
}{
Although these figures seem high, when integrating the quantity
$\lambda F_\lambda $ across the extrapolated spectrum between
0.84--2.30~\microns, ignoring telluric regions in this integral, the difference
is only $-3.1$\%. This difference in the integral gives perhaps a better figure
of merit when making corrections to photometry than looking at only the median
percent-difference in the flux.
}

% 21fxy
% Shallow-silicon, 95.45%
% Median past 1.5u : 3.970
% Median in 0.9u < lambda < 1.5u : -5.562
% Integral F * lam % difference : -2.863
% Si II
% Median past 1.5u : 4.827
% Median in 0.9u < lambda < 1.5u : -5.235
% Integral F * lam % difference : 1.993

In the top-right panel of \autoref{fig:single_combined}, a near-max spectrum of
SN~2021fxy \citep{DerKacy_etal_2023fxy} is shown alongside its NIR prediction.
The observation is, similar to the SN~2011fe spectrum from
\citep{Mazzali_etal_2014_11fe}, a merge of an optical spectrum around $+0$
days and a NIR one around $+1$ days. This process of merging normalizes the NIR
spectrum to the optical spectrum by the integrated flux within 0.84--0.88
\microns, and performs
a weighted average to get the new merged flux for that region. SN~2021fxy was
determined by GMMs to be a SS \sn with a probability of 95.5\%. The prediction
here is quite similar to that of SN~2011fe, however all merits of fit discussed
for SN~2011fe were more true to the observation for SN~2021fxy, even with a
noisier spectrum, aside from a poorer fit to NIR Ca~II around 0.8~\microns. In
the 0.9--1.5~\microns region, the median difference in flux was $-5.6$\%, and
past 1.5~\microns the median is $+4.0$\%. Integrating the observed and
the predicted spectra in the same way as before shows a difference of $-2.9$\%.
It seems that although the eigenvectors shown in \autoref{fig:eigenvectors}
appear to show substantial variation of \SiII features, shallower \SiII
lines that are projected onto these eigenvectors do not lead to a less accurate
prediction.

% 21aefx
% Core-normal, 57.63%
% Median past 1.5u : 7.791
% Median in 0.9u < lambda < 1.5u : -13.568
% Integral F * lam % difference : -2.717
% Si II
% Median past 1.5u : 5.030
% Median in 0.9u < lambda < 1.5u : -14.475
% Integral F * lam % difference : -0.349

The bottom-left panel of \autoref{fig:single_combined} shows a
near-maximum spectrum of SN~2021aefx
\citep{Burns_etal_POISE,Hosseinzadeh_etal_2022_21aefx} as well as its
extrapolation. The observed spectrum is merged in the same way as SN~2021fxy
with an optical spectrum at $+3$ days and a NIR spectrum at $-1$ days.
SN~2021aefx is classified with GMMs as a CN with a probability of only 57.6\%
and therefore may exhibit some atypical spectroscopic behavior relative to most
CN \sne. Within 0.50--0.84~\microns, the fit to the observation including \SiII
is good except for the noisier region around 0.80~\microns where there is more
fringing of the original optical spectrum. As such, the fit to NIR Ca~II is
poor. Still, the NIR prediction of this spectrum performs reasonably, with a
median percent difference in flux of $+7.8$\% until 1.5~\microns where this
difference increases, reaching a median of $-13.6$\%. SN~2021aefx continues to
compare well to the observation with a difference in integrated flux of
$-2.7$\%.

% 22hrs
% Broad-line, 99.98%
% Median past 1.5u : -21.357
% Median in 0.9u < lambda < 1.5u : -17.312
% Integral F * lam % difference : -17.745
% Si II
% Median past 1.5u : -19.005
% Median in 0.9u < lambda < 1.5u : -22.301
% Integral F * lam % difference : -18.531

Finally, the bottom-right panel of \autoref{fig:single_combined}
\deleted{
,
}
shows a
100\%-likely BL \sn (rounded up from a GMM calculation of 99.98\% probability),
SN~2022hrs \citep[][P. Brown et al., in prep.]{Burns_etal_POISE}. The
observation shown here is made from merging an optical spectrum at $-1$ days and
a NIR spectrum at $+6$ days. This \sn provides an interesting
outlook as no BL \sne exist in the maximum-light training data set used to make
this prediction. This possibly explains the relatively poor fit to the
observation in the 0.50--0.84~\microns region, and especially when reproducing
\Rline. It seems that a linear combination of PCs that is able to describe the
depth of \Rline alongside other features not obtainable without more BL data
included in the training sample. There are also larger
uncertainties involved with this spectrum. The most glaring difference between
this \sne and the other three examples is that there seems to be a $\sim$20\%
near-constant negative offset between the prediction and observation after
0.84~\microns. This again may be due to the eigenvector training sample not
including BL \sne, and therefore a BL such as SN~2022hrs is not represented
well enough to accurately reconstruct its spectrum in either the optical or the
NIR. In fact, we see this later in \autoref{sec:color-matching} with SN~2014J,
a CN \sn with high \vsi, similar to BLs.
This offset may also be due to the seven-day difference in observation time
between the optical and NIR spectrum used to create the observation shown.
With this offset seen for SN~2022hrs, the difference in integrated flux between
the prediction and observation is $-17.7$\%, substantially greater than the
other three example \sne.

\added{
\begin{figure*}[ht]
    \centering
    \includegraphics[width=\textwidth]{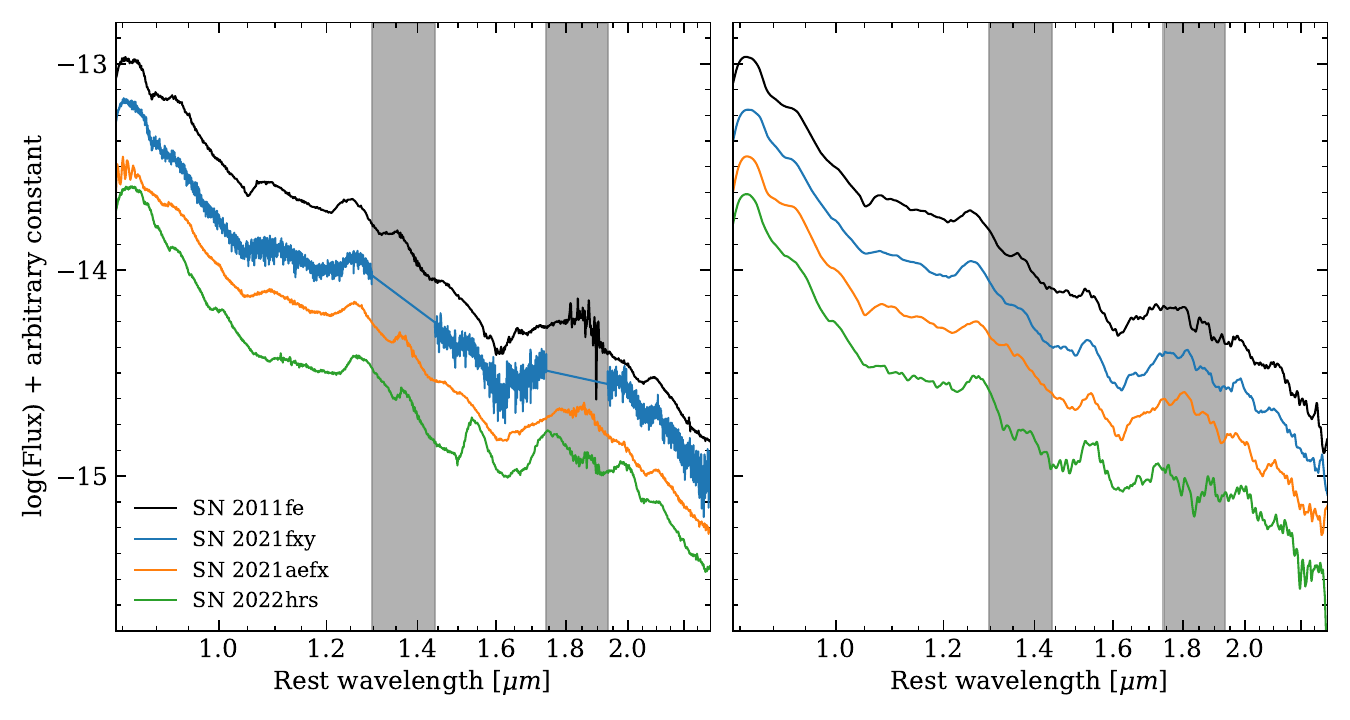}
    \caption{A direct comparison of the four near-maximum-light observations
             from \autoref{fig:single_combined} (left panel) and their
             corresponding extrapolations (right panel) in the NIR. The
             extrapolations appear to capture many of the distinctions shown
             between the observed spectra, except for the Fe~II feature near
             1.50~\microns.}
    \label{fig:compare_predictions}
\end{figure*}

Although \autoref{fig:single_combined} demonstrates a good comparison between
predictions and their respective observed spectra, it is difficult to determine
any significant difference between the predictions, and whether or not the
characteristic differences between \sn subtypes are being represented well by
the predictions. For this reason, we also provide a direct comparison of the
predictions themselves in \autoref{fig:compare_predictions}. The left panel
shows the observed spectra from each \sn in \autoref{fig:single_combined} past
0.84~\microns. The spectrum in the telluric regions of SN~2021fxy was omitted
for visual clarity due to the high amount of noise. In the right panel, the
corresponding predictions (red lines from \autoref{fig:single_combined}) are
plotted. Each spectrum was normalized by the integrated flux of the
0.84--2.30~\microns region, ignoring the telluric regions in the grey shaded
region, with SN~2011fe as the reference. The spectra are then equally spaced
from each other in the figure.

Here it is beneficial to summarize the Branch groups to which each \sn here
belongs: SN~2011fe and SN~2021aefx are classified CNs, SN~2021fxy is a SS, and
SN~2022hrs is a BL \sn. Keeping this in mind, the predictions seem to do very
well at the feature level in capturing the differences between these \sne.
Using line identifications by \citet{Hsiao_etal_2019}, Mg~II around
0.92~\microns, 1.09~\microns is conserved in the prediction, being stronger
for the CNs, slightly weaker for the SS, and weakest for the BL \sn. However,
the predicted Mg~II at 1.09~\microns appears much more narrow than the
observation for SN~2021aefx. Mg~II near 2.10~\microns appears strong for the BL
\sn as well, and the prediction captures the observations well for this
feature, including the feature shape. Si~III near 1.25~\microns is similar for
each \sn in both the observation and the prediction. Using line identifications
displayed by \citet{Friesen_etal_2014}, the Co~II feature around 1.65~\microns
is also differentiated well in the prediction for each subtype, including the
feature shape. The biggest departure from the observation that the prediction
contains is the Fe~II feature around 1.50~\microns. Fe~II here is much more
shallow for CNs, which is not expressed by the predictions. In general, though,
the feature behavior between \sne of different Branch groups appears to be
captured well by these predictions.
}

\subsection{Time-Dependent Predictions}

\begin{figure*}[ht]
    \centering
    \includegraphics[width=\textwidth]{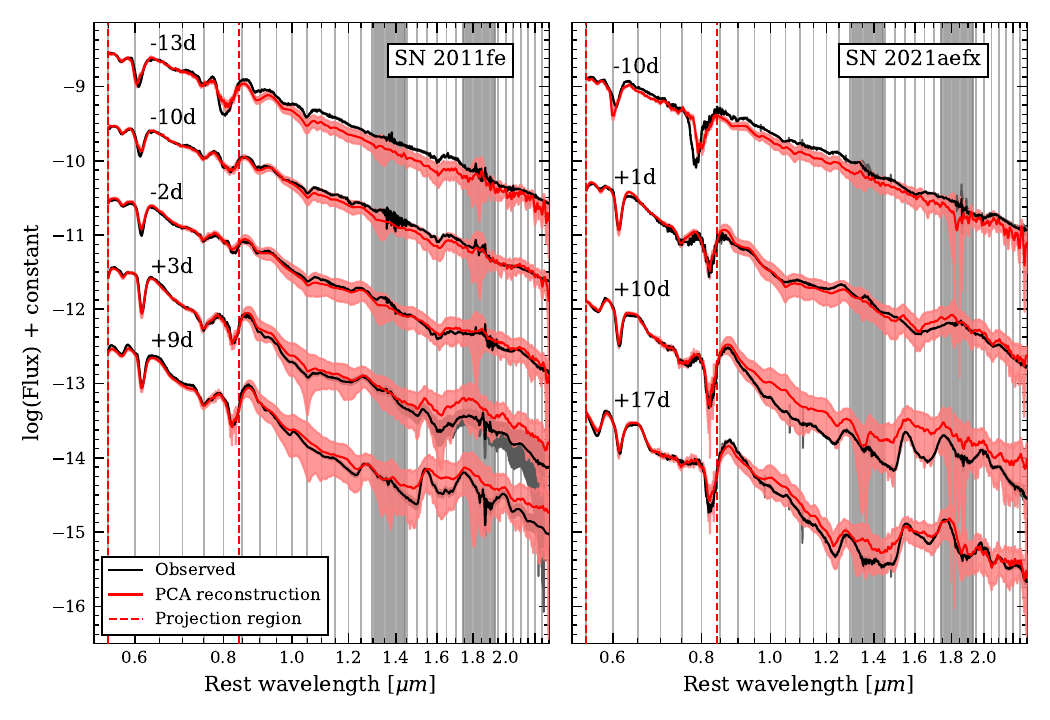}
    \caption{Extrapolation for a time series of SN~2011fe and SN~2021aefx. All
             formatting and notation is the same as in
             \autoref{fig:single_combined}.}
    \label{fig:time_series}
\end{figure*}

We showed in \autoref{sec:results-max-light} that the extrapolation process
performs
fairly well for spectroscopically normal \sne near maximum light. However, when
extrapolating at other phases, the results and performance of this technique
may lead to less certain predictions, especially due to decreasing training
sample sizes. As expected from the sample sizes given in
\autoref{fig:sample_times_true}, at later times the available number of
training spectra decreases significantly as there are not as many concurrent
optical and NIR observations in the CSP I+II data set.
Interpolation at these later times is often not possible, as either the total
number of observations of each individual \sn are too few, or the observations
were too
far apart in time to give reasonable interpolations. However, up to around 20
days past maximum light, the available sample size is still around 20 \sne
--- twice the number of eigenvectors we use to model spectra, which may
generate reasonable extrapolations of spectra observed at these phases.

In \autoref{fig:time_series} we show the same prediction process as in
\autoref{sec:results-max-light} on a time series of early spectra of SN~2011fe
\citep{Mazzali_etal_2014_11fe} and SN~2021aefx
\citep{Burns_etal_POISE,Hosseinzadeh_etal_2022_21aefx}, as they have good
concurrent observations in both the optical and
NIR surrounding maximum light. Both \sne yield similar results. At earlier
times, most feature behavior is captured, aside from Ca~II around
0.80~\microns.
The prediction for SN~2011fe also introduces a small constant offset from the
prediction across the NIR that is not exhibited by SN~2021aefx. This may be
due to differences in the flux calibrations between the optical and NIR
spectra, which cannot be corrected for in the merging process due to a lack of
photometric data for the entire data set (see \autoref{sec:color-matching}).

At later times, the general trends of the flux of both \sne are captured;
however, much of the feature information beyond 1.5~\microns appears to be lost,
and the uncertainties are much larger. This could mean that there is more
variation in this region than can be explained with the limited sample
size available.

\subsection{Prediction Using \SiII}

For consistency we used the 0.50-0.84~\microns region to project onto the
eigenvector basis in order to make predictions. However, this region itself is
a variable and may be changed. Increasing this range is impractical as, again,
using information blueward of 0.50~\microns introduces much of the UV-optical
variation that is difficult to explain using a limited training sample.
Instead, this section focuses on the outcome of narrowing this region to a more
meaningful one.

\begin{figure*}[ht]
    \centering
    \includegraphics[width=\textwidth]{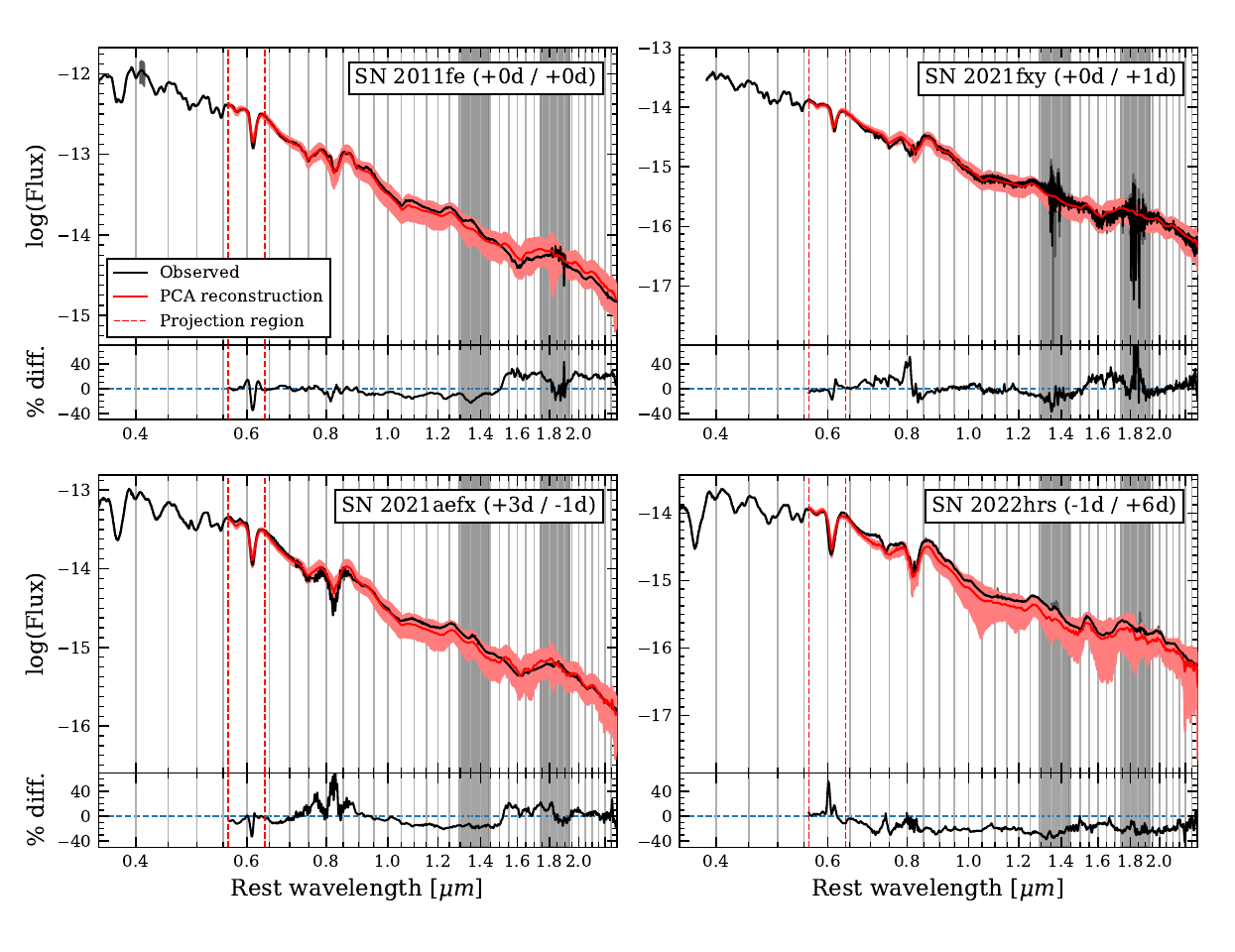}
    \caption{This shows the same spectra as in \autoref{fig:single_combined},
             however the predictions are now made with a narrower optical
             region dominated mostly by \Rline and \Bline.
             \added{
             All formatting and notation is the same as in
             \autoref{fig:single_combined}.
             }}
    \label{fig:single_combined_SiOnly}
\end{figure*}

It is clear that the \Rline and \Bline features in the optical are important
observables in the diversity of \sne, as they make up the basis of the Branch
classification system, which has been shown to be yield a statistically
significant set of groups with distinct characteristics
\citep{Burrow_etal_2020}. \autoref{fig:single_combined_SiOnly} shows the result
of projecting only the 0.55--0.64~\microns region consisting of these two \SiII
lines onto the eigenvector basis to extrapolate into the NIR up to
2.3~\microns. This was done for the same \sne as in
\autoref{sec:results-max-light}, and the figures show the same information as
shown in \autoref{fig:single_combined}; the only change is the region used to
make the prediction, shown as red dashed lines. It is important to reiterate
that the only information provided to the maximum-light model
is this narrow \SiII window --- aside from the mean flux value of the
0.50-0.84~\microns region required to normalize the spectrum in the same way as
the training data.

The predictions shown in \autoref{fig:single_combined_SiOnly} are quite
surprising, given the small amount of information provided. Comparing SN~2011fe
and SN~2021fxy in \autoref{fig:single_combined} and
\autoref{fig:single_combined_SiOnly}, the predictions are quite similar. For
each \sn, both cases seem to predict the general flux behavior of the
observation fairly well. The uncertainties shown are nearly the same across
each spectrum between the two cases, except they are much improved in the case
of SN~2022hrs where spikes of larger uncertainties are shown at some
wavelengths past 1.5~\microns. The most notable difference between these two
cases is the inability to reproduce the NIR Ca~II triplet well. This seems to
suggest that Ca~II behavior is not strongly dependent on \Rline and \Bline near
maximum light.

% 11fe
% Integral F * lam % difference : -3.083
% 21fxy
% Integral F * lam % difference : -2.863
% 21aefx
% Integral F * lam % difference : -2.717
% 22hrs
% Integral F * lam % difference : -17.745

% Si only:

% 11fe
% Integral F * lam % difference : -2.366
% 21fxy
% Integral F * lam % difference : 1.993
% 21aefx
% Integral F * lam % difference : -0.349
% 22hrs
% Integral F * lam % difference : -18.531

The integrated flux differences from the observations in this case are $-2.4$\%
for SN~2011fe, $+2.0$\% for SN~2021fxy, $-0.4$\% for SN~2021aefx, and $-18.5$\% for
SN~2022hrs. This means that the prediction was marginally improved for
SN~2011fe, SN~2021fxy, and SN~2021aefx, and only weakened for SN~2022hrs, which
is likely because it is a BL, unlike the \sne in the training set. Given
just these few test spectra shown here, no conclusion is drawn regarding which
case performs better. However, the similarity in results between the two cases
indicates that \SiII may be the only requirement for a reasonable
prediction in the NIR in general.

\section{Discussion}
\label{sec:discussion}

\subsection{PC Relation to Branch Groups}
\label{sec:pcs_branch}

Because the NIR spectrum seems well-inferred by projecting only the two \SiII
features onto the eigenvectors, one may think that the eigenvectors themselves
have some relation to the Branch group classification scheme. Here we look at
how these eigenvectors relate to the Branch groups to explore the possibility
that each individual Branch group has unique spectroscopic features that
exhibit a relationship with features in the NIR.

\begin{figure*}[ht]
    \centering
    \includegraphics[width=\textwidth]{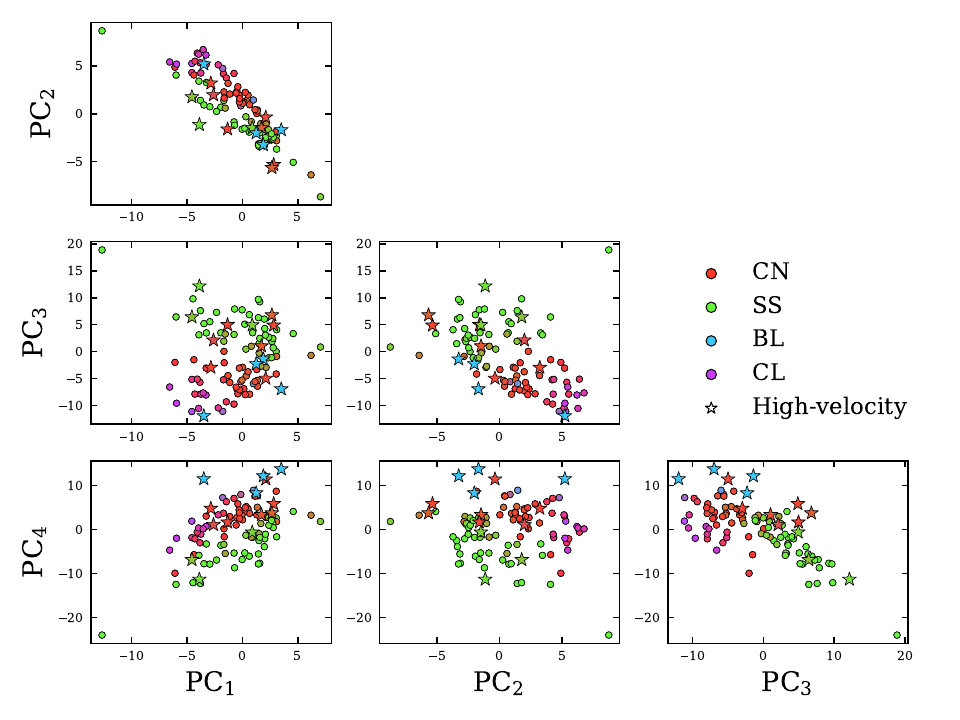}
    \caption{The projections of spectra in the optical data set
             onto the first four maximum-light PCs in \autoref{fig:eigenvectors}
             only in the 0.55~\microns $< \lambda <$ 0.66~\microns region.
             Points are colored for the probability of Branch group
             classification in the same way as in \cite{Burrow_etal_2020}
             using a 3-D GMM. Points shown as stars are high-velocity \sne
             exhibiting \vsi $>$ 12,500~\kmps
             \added{
             at maximum light
             }.}
    \label{fig:pcs_branch}
\end{figure*}

We project 101 near-maximum-light spectra (within $\pm$ 5 days of maximum
light) available from
the optical CSP~I~\&~II data set onto the maximum-light eigenvectors
shown in \autoref{fig:eigenvectors}. However, we only project the region
between 0.55--0.66~\microns in order to account for the extent of \Rline and
\Bline on
the eigenvectors. These values --- denoted as $\text{PC}_1$, $\text{PC}_2$,
etc., corresponding to the projection onto the respective eigenvectors --- are
then plotted against each other in the correlation matrix shown in
\autoref{fig:pcs_branch}. The pEWs of \Rline and \Bline as well as \Rline
expansion velocity \vsi of each spectrum were calculated using Spextractor.
By using a 3-D Gaussian mixture model from \cite{Burrow_etal_2020}, the
Branch group that each \sn is assigned to lie within has been colored in a
similar fashion to \cite{Burrow_etal_2020}. In addition, \sne exhibiting higher
velocities with \vsi $>$ 12,500~\kmps are symbolized with stars in
\autoref{fig:pcs_branch}. Many of these high-velocity (HV) \sne are classified
as CNs or BLs, which is seen in \cite{Burrow_etal_2020}.

It is immediately apparent that a few subspaces, for example $\text{PC}_1$ vs.
$\text{PC}_3$ or $\text{PC}_2$ vs. $\text{PC}_4$, seem to show CN, SS, and CL
\sne lying in distinct regions, similar to a classic Branch diagram. There are
some exceptions to this, for example in the $\text{PC}_1$ vs. $\text{PC}_3$
panel, several CNs lie in the region dominated by SS \sne. This illustrates
that there is not a one-to-one relationship between these projections and the
measured pEWs of \Rline and \Bline, and therefore this is not a recreation or
rotation of the classic Branch diagram. This is further explored in the
discussion
of \autoref{tab:pc_vs_properties} later in this subsection. This suggests that
there may be some intrinsic differences between these CNs lying within the
SS-dominated region in this subspace and the other more concentrated CNs.
Many of these CNs within the SS region exhibit higher velocities; however, this
likely cannot be the only characteristic that sets these \sne apart from other
CNs, because other HV CNs are seen in the concentrated region as well. Finally,
note that BL \sne are not as constrained to their own regions in
\autoref{fig:pcs_branch}. This could be due to the unfortunate lack of BL
representation in the training sample used to calculate these eigenvectors.
Because of this, it is difficult to arrive at concrete conclusions for BL \sne
from this figure.

\begin{figure*}[ht]
    \centering
    \includegraphics[width=\textwidth]{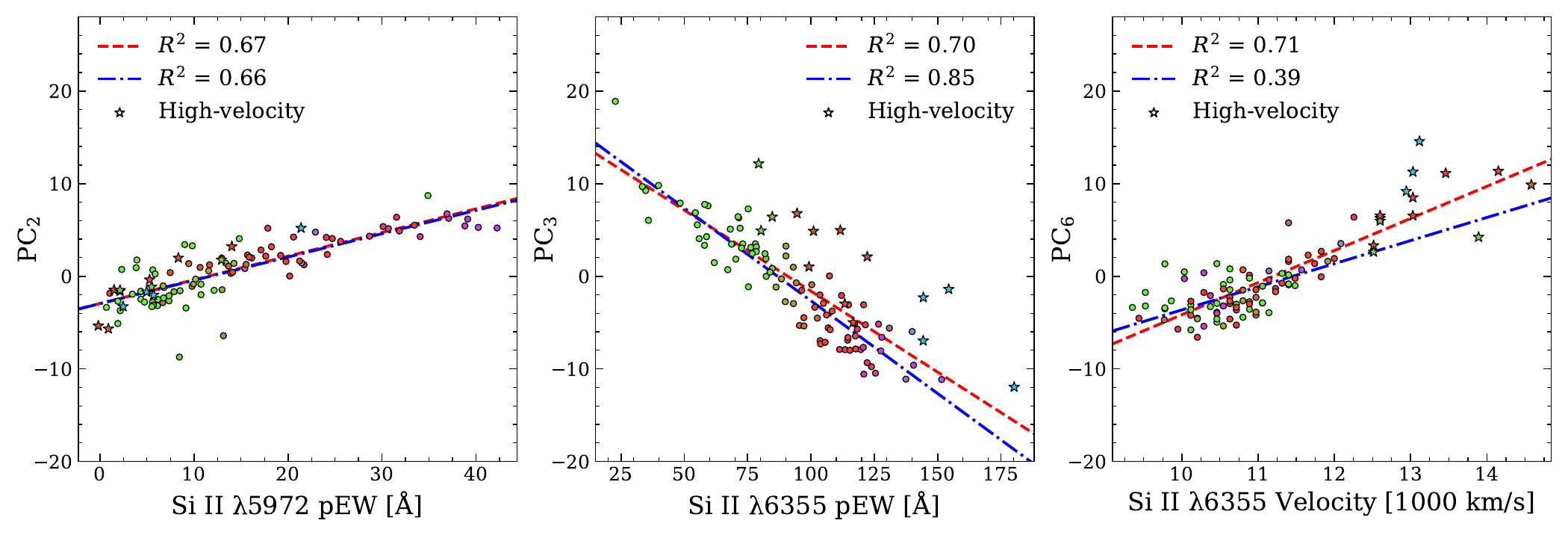}
    \caption{$\text{PC}_2$, $\text{PC}_3$, and $\text{PC}_6$ plotted against
             pEW(\Bline), pEW(\Rline), and \vsi, respectively. Colors here are
             the same 3-D GMM colors shown in \autoref{fig:pcs_branch}, and
             points shown as stars indicate the same HV \sne. Lines here
             indicate linear regression lines for all the points (in red,
             dashed) and for all points except HV \sne (in blue, dash-dotted).}
    \label{fig:pc_vs_property_condensed}
\end{figure*}

\begin{deluxetable*}{l|ccc}
    \tablecaption{Summary of $R^2$ values of all $\text{PC}_i$ linear
                  correlations with various spectroscopic properties. Values in
                  parentheses indicate the $R^2$ value of the linear regression
                  excluding HV \sne.
        \label{tab:pc_vs_properties}}
    \tablewidth{0pt}
    \tablehead{
        \multicolumn{1}{c|}{Projection} & \multicolumn{3}{c}{$R^2$} \\
        \multicolumn{1}{c|}{} & \colhead{[pEW(\Bline)]} & \colhead{[pEW(\Rline)]} & \colhead{[\vsi]}}
    \decimals
    \startdata
        $\text{PC}_1$ & 0.42 (0.44) & 0.02 (0.03) & 0.03 (0.06) \\
        $\text{PC}_2$ & 0.67 (0.66) & 0.19 (0.29) & 0.03 (0.00) \\
        $\text{PC}_3$ & 0.34 (0.35) & 0.70 (0.85) & 0.00 (0.08) \\
        $\text{PC}_4$ & 0.00 (0.00) & 0.49 (0.43) & 0.23 (0.35) \\
        $\text{PC}_5$ & 0.51 (0.51) & 0.01 (0.03) & 0.42 (0.20) \\
        $\text{PC}_6$ & 0.04 (0.00) & 0.17 (0.06) & 0.71 (0.39) \\
        $\text{PC}_7$ & 0.09 (0.11) & 0.61 (0.62) & 0.01 (0.05) \\
        $\text{PC}_8$ & 0.03 (0.03) & 0.46 (0.52) & 0.02 (0.16) \\
    \enddata
    % \tablecomments{}
\end{deluxetable*}

Investigating the $\text{PC}_i$ subspaces further,
\autoref{fig:pc_vs_property_condensed} plots the projections of \SiII onto PCs
2, 3, and 6 against pEW(\Bline), pEW(\Rline), and \vsi, respectively. Each
point is colored by Branch group classification in the same way as
\autoref{fig:pcs_branch}. Linear regression lines are shown for each
panel; the red line shows the regression for all points, and the blue line
shows the same for all points except high-velocity \sne. The $R^2$ coefficient
of determination for each regression line is shown in the legend of each panel.
A summary of the $R^2$ values for each projection up to $\text{PC}_8$ against
all three spectroscopic properties is given in \autoref{tab:pc_vs_properties},
where $R^2$ values from fits excluding HV \sne are given in parentheses. PCs 2,
3, and 6 were chosen to be displayed because they exhibit the highest $R^2$ for
correlations with the three properties for both the general and non-HV cases.
Although a low value of $R^2$ does not rule out a non-linear correlation, other
values of $\text{PC}_i$ are not displayed here because there is no clear visual
trend in them, or the spread of the points is too large. This
leads to the low values of $R^2$ for the majority of projections seen in
\autoref{tab:pc_vs_properties}.

\replaced{
It is clear, though, that the
}{
The
}
highlight of this analysis is obtained by
observing how $\text{PC}_3$ relates to the pEW(\Rline) shown in the middle
panel of \autoref{fig:pc_vs_property_condensed}. There is a significant spread
of \sne around the regression line, however when ignoring the clearly distinct
HV \sne in this panel, the spread is greatly reduced, leading to a
significantly improved $R^2$, which is indicative of a stronger linear
correlation between $\text{PC}_3$ and pEW(\Rline). The effect of HV \sne being
outliers from the trend line is also not dependent on Branch group
classification --- i.e., these outlying points do not only correspond to BL
\sne. It appears that the measurement of $\text{PC}_3$ seems to be accounting
for HV behavior in a way that the classic measurement of pEW(\Rline) cannot.

In addition, we find a mild linear correlation between $\text{PC}_2$ and
pEW(\Bline), which itself has been shown to exhibit a connection with \sbv
\citep[see, e.g.,][]{Burrow_etal_2020}. However, we show and discuss in
\autoref{sec:sbv} that more information than $\text{PC}_2$ is needed to extract
a estimate of \sbv from a spectrum using these projections. The slope estimator
of each linear regression shown in \autoref{fig:pc_vs_property_condensed} has a
two-sided P-value of much less than 0.001, heavily suggesting that there is at
least some dependency of these three $\text{PC}_i$ and the respective
spectroscopic quantities shown. Visually, though, the wide spread of measured
$\text{PC}_2$ seems comparable to the change in the trend of $\text{PC}_2$ as
pEW(\Bline) changes and therefore, this linear relationship seems
untrustworthy as an estimator. This also seems true for $\text{PC}_6$ versus
\vsi, as removing HV \sne in this panel heavily diminishes the quality of this
linear correlation, which is suggested by the much lower $R^2$ statistic.

These projections may offer an alternative method of classifying \sne
that does not rely on the assumption of a pseudo-continuum that can lead to
inconsistent measurements. It would be interesting to perform further testing
with these $\text{PC}_i$ measurements to decide their candidacy as
classification criteria that may offer solutions to some known discrepancies in
\sne classification. For example, \cite{Burrow_etal_2020} show that several
CNs exhibit high velocities that would be expected of BLs, given the
relationship between the velocity of the blueshifted absorption part of a
P~Cygni profile and it's equivalent width. Measuring $\text{PC}_i$ values
instead of pEWs could reclassify these HV CNs to be similar to \sne with
broader \Rline features.

\subsection{PC Relation to $s_{BV}$}
\label{sec:sbv}

\begin{figure*}[ht]
    \centering
    \includegraphics[width=\textwidth]{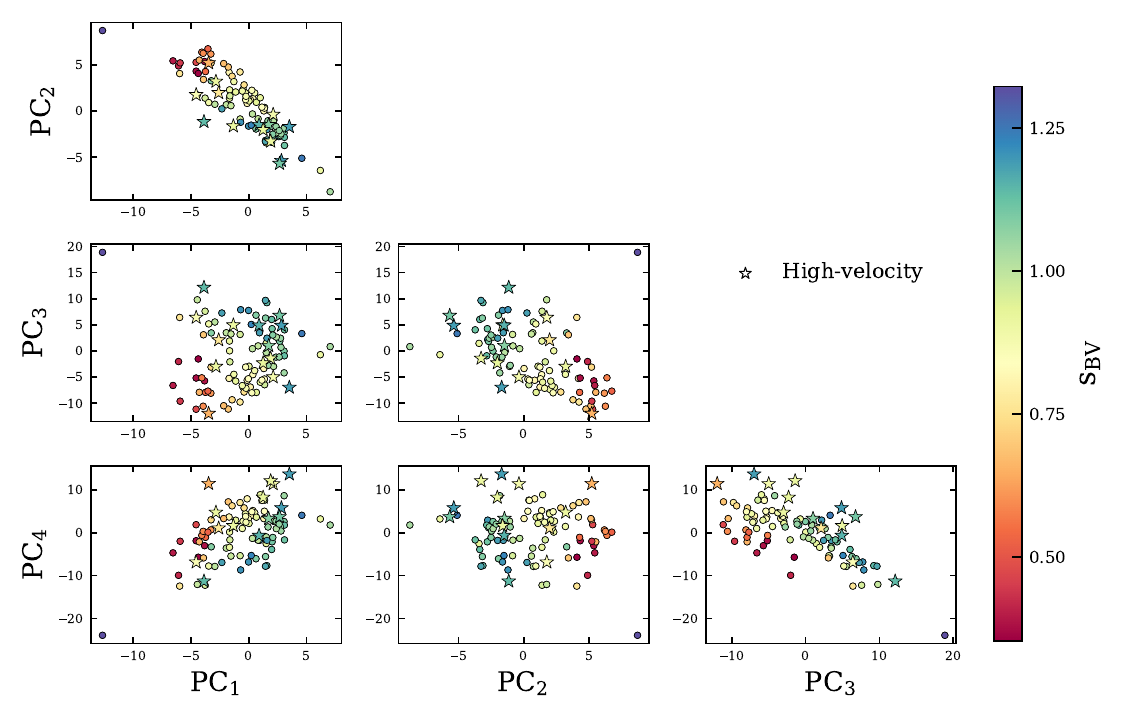}
    \caption{The same projections as in \autoref{fig:pcs_branch}, however now
             colored for \sbv determined by SNooPy. It seems visually plausible
             that \sbv may be estimated here as a function of two or more PCs.}
    \label{fig:pcs_sbv}
\end{figure*}

\added{
\begin{figure}[ht]
    \centering
    \includegraphics[width=\linewidth]{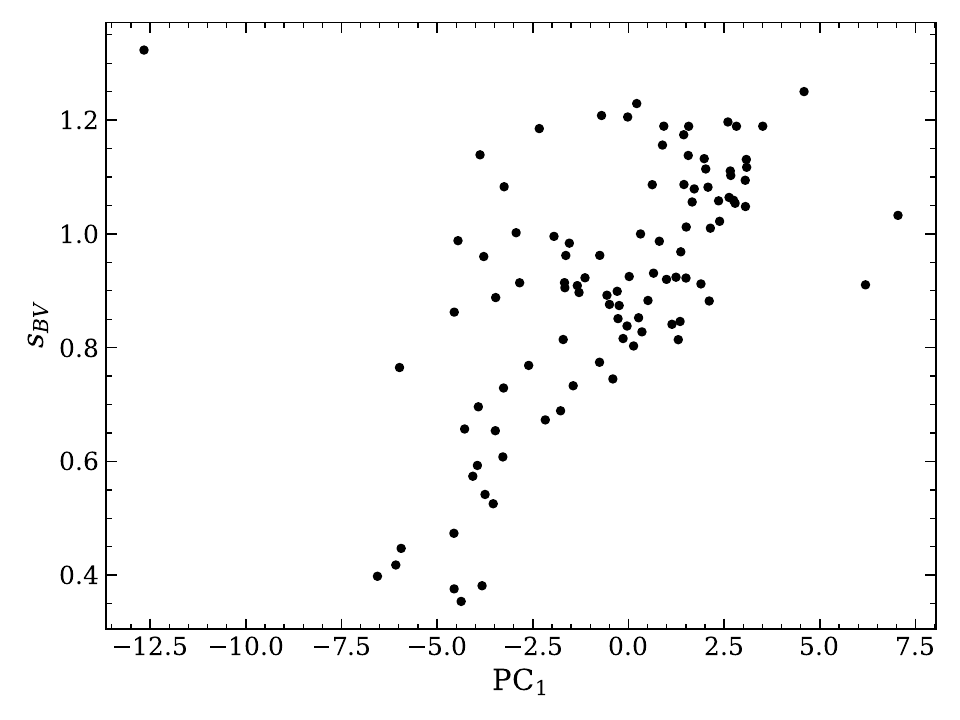}
    \caption{A direct view of $\text{PC}_1$, from \autoref{fig:pcs_sbv},
             versus \sbv. This shows the lack of correlation between the two,
             showing that more information than $\text{PC}_1$ is needed to
             estimate \sbv.}
    \label{fig:pc1_sbv}
\end{figure}
}
  
Similar to \citet{Lu_etal_2023}, it is interesting to see how these PCs may be
related to light-curve parameterization such as the color-stretch
parameter \sbv. This relation may play an important role in the connection
between observed photometry and the underlying SED of \sne. In
\autoref{fig:pcs_sbv}, the exact same PC projections as those shown in
\autoref{fig:pcs_branch} are displayed, now colored for the \sbv value
of each \sn calculated using SNooPy. There appears to be a multi-dimensional
trend between \sbv and many of the PC values. In other words, by projecting the
\SiII region of a spectrum onto the maximum-light eigenvectors, this would
produce values that should allow one to predict \sbv given only spectroscopic
information. For example, just using $\text{PC}_1$ and $\text{PC}_2$, it may be
seen that \sne with high values of $\text{PC}_1$ and low values of
$\text{PC}_2$ tend to have higher \sbv, and vice versa.

\replaced{
It is clear by looking at the $\text{PC}_1$ column that $\text{PC}_1$ does not
directly correlate with \sbv by itself.
}{
Looking at the $\text{PC}_1$ column, it appears that $\text{PC}_1$ does not
directly correlate with \sbv by itself.
}
\added{
This is verified by \autoref{fig:pc1_sbv}, which plots $\text{PC}_1$ against
\sbv, showing the large amount of spread in the relationship between the two
quantities. If one were to use this relationship with a measured value of
$\text{PC}_1$, it would not be possible to estimate an accurate \sbv without
large uncertainty.
}
This
\replaced{
hints
}{
heavily suggests
}
that the first PC shown in \autoref{fig:eigenvectors} does not describe the
same variation of \sne within the training sample that \sbv explains.

\begin{figure}[ht]
    \centering
    \includegraphics[width=\linewidth]{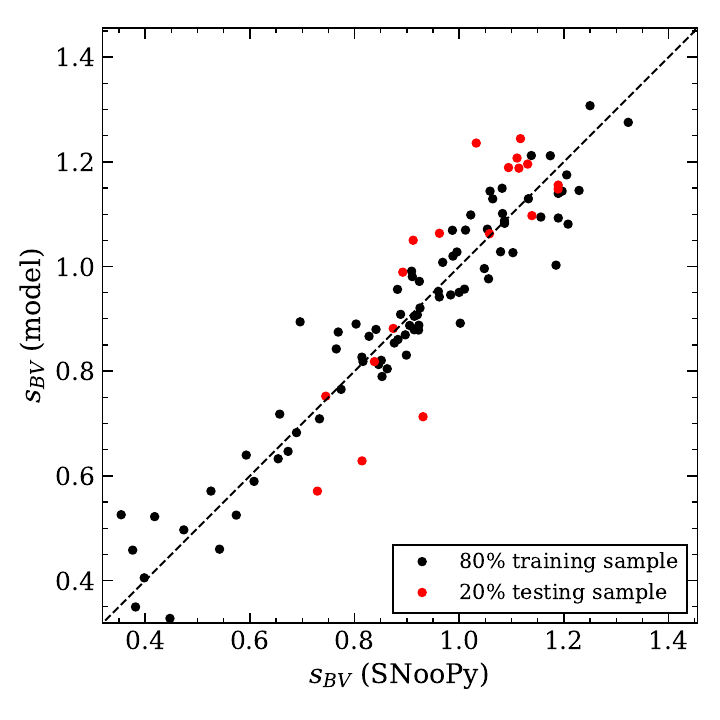}
    \caption{On the x-axis, \sbv calculated by SNooPy is shown for the same
             \sne that produce the points in \autoref{fig:pcs_sbv}. On the
             y-axis, \sbv is shown, calculated from a linear model described
             by \autoref{eq:sbv}, where the $a_n$ coefficients are
             determined via a least-squares fit to a random 80\% of the
             available spectra representing black points. The red points are
             the remaining 20\%, illustrating this \sbv model fits well with the
             majority of values given by SNooPy.}
    \label{fig:model_sbv}
\end{figure}

Examining each subplot in \autoref{fig:pcs_sbv}, the regression of \sbv
visually appears to define a direction in each subspace, suggesting that \sbv
may be determined by the projections. This idea of predicting \sbv may be put
to the test with a simple linear model. This model may be a function of
any number of PC values, i.e.,
\begin{equation}
\label{eq:sbv}
s_\text{BV} = a_0 + \sum_n^N a_n \text{PC}_n,
\end{equation}
where $a_n$ are coefficients for each projection $\text{PC}_n$ up to the
projection $\text{PC}_N$. After some testing, we find that the sum of
least-squares are reduced as more projections are added to this model (and
thus $N$ increases). For this example we extend this model up to $N = 8$, and
therefore the linear model will contain nine coefficients to be fit. These
coefficients are determined by fitting to the data in \autoref{fig:pcs_sbv}
using least-squares fitting.

To simulate a test of this procedure, we take a random 80\% (82 points) of
the \sne in \autoref{fig:pcs_sbv} to fit the coefficients of the model, then
use this model to estimate \sbv of the same \sne. These values are displayed in
\autoref{fig:model_sbv} as black points. The remaining points colored red in
\autoref{fig:model_sbv}
are the \sbv model estimates of the remaining 19 \sne, from which no
information was provided to the model. These \sbv values are then compared with
those obtained by the light-curve-fitting tool SNooPy.

We see that the vast majority of these \sne adhere to this model
within a spread of about $\pm 0.1$ units of \sbv for all values of measured
\sbv. The testing sample seems to be commensurate with the training sample for
all \sbv, aside from three outlying test points around $0.7 <$ \sbv $< 1.0$,
although they do not lie too far outside this trend. In general, for such a
simple linear model and method (using of a small region of the spectrum),
this fit to \sbv values measured with SNooPy is surprisingly good. One could
possibly use a more complex regression technique or test other regions in
wavelength space to produce better fits
\added{
(at the risk of over-fitting)
};
however, the purpose of this is to
show the simplicity of the correlation of \SiII properties with \sbv. In
addition, for future studies, it could be interesting to see if any outliers
to this model exhibit larger residuals in its NIR light-curve as determined by
SNooPy.
If an observed spectrum yields a substantially different value of \sbv for a
supernova using these PCs compared to using the observed photometry, this could
indicate intrinsic differences in its spectrum which could be seen in its
NIR light-curves.

\subsection{Color-Matching with Broad-Band Photometry}
\label{sec:color-matching}

As discussed in \autoref{sec:data}, neither training spectra nor any spectra
shown in \autoref{sec:results} have been color-matched to the broad-band
photometry. As such, the
differences in the flux calibrations between optical and NIR
spectra may result in some training spectra having improperly scaled NIR flux.
In other words, the photometric color of the predicted spectrum will be
incorrect. This may lead to substantial variation in the training data in the
NIR. However, if this variation is in any way systematic, the eigenvectors may
describe this variation and be able to correct for it, albeit not perfectly.
It is not possible to know if this hypothesis is true without observed
photometry for all of our training \sne; however, we can still make this
assumption. This
assumption may be supported by the first PC in \autoref{fig:eigenvectors},
which seems to describe a broad correction that accounts for around 58\% of the
variation (see \autoref{fig:explained_variance}) in the training sample. We
\replaced{
see
}{
discuss
}
in \autoref{sec:sbv} that $\text{PC}_1$ does not correlate by itself with \sbv,
and therefore the first maximum-light eigenvector likely does not describe
variation of a stretch-like property of \sn light-curves. Instead, it may be
correcting for differences in the flux calibration between optical and NIR
spectra --- if so, the final NIR prediction should be on the appropriate
scale of the optical spectrum.

\begin{figure}[ht]
    \centering
    \includegraphics[width=\linewidth]{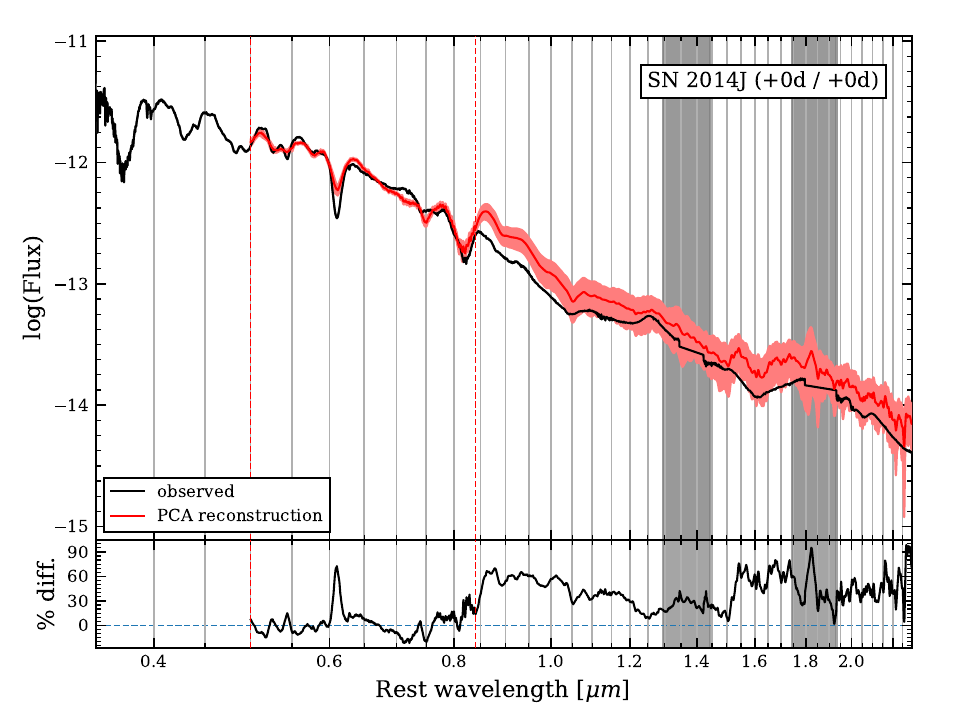}
    \caption{Maximum-light extrapolation for SN~2014J. All formatting is the
             same as that in \autoref{fig:single_combined}.
             \added{
             Although SN~2014J is classified as a CN \sn, a constant offset is
             prominently seen between the observation and prediction. This may
             suggest that photometric corrections to spectra are necessary to
             strengthen the performance of these extrapolations, which
             requires more photometric observations in both the optical and
             NIR.
             }
            }
    \label{fig:14J_formal}
\end{figure}

% 14J
% Broad-line, 78.82% (Default spex)
% Core-normal, 67.64% (Manual range w/ left side up to the next max)
% Median past 1.5u : 46.721
% Median in 0.9u < lambda < 1.5u : 32.016
% Integral F * lam % difference : 43.186

An illustration of this problem is shown in \autoref{fig:14J_formal}, which
shows a maximum-light prediction of SN~2014J using a merged optical-NIR
spectrum \citep{Galbany_etal_2016_14J,Srivastav_etal_2016_14J} in the
same way as those in \autoref{sec:results-max-light}.
\added{
SN~2014J is known to be highly reddened
\citep{Amanullah_etal_2014,Foley_etal_2014}. We assume host-galaxy color-excess
values of $E(B-V)_{host} = 1.24$ mag and $R_V = 1.44$ \citep{Ashall_etal_2014}
to correct the spectrum for extinction using the CCM law \citep{CCM_1989},
similar to spectra in the PC training data.
}
Even though the merging
and prediction process is the same as the other four \sne that yields good
fits, the case of SN~2014J is unexpectedly inadequate. Inside the
0.50--0.84~\microns region which was projected onto the eigenvectors, the PCA
reconstruction fails to fit the observation for most features in this region,
and this is especially true for \SiII. It seems as though SN~2014J does not
exhibit the same behavior as the training sample and therefore could be an
interesting case study on its own. It is worth noting that, although we
classify SN~2014J as a CN with a probability of 67.6\%, it is also on the
faster end of CNs with \vsi of $12.0 \times 10^3$ \kmps and could nearly be
considered a BL \sn by traditional classification with only pEWs. However, the
GMM probability of belonging to the BL group is 13.5\%. It was shown in
\autoref{sec:results-max-light} that this prediction process is not suitable
for BLs such as SN~2022hrs, as a constant offset from the prediction was
produced. Perhaps the same underlying reasoning is being displayed for the case
of SN~2014J. This shows that more observations of BL and faster \sne are needed
for a complete understanding of the diversity of \sne.

Similar to SN~2022hrs, SN~2014J shows a large and nearly constant offset
seen between the prediction and the observation after 0.84~\microns. Because
the observations shown in black are simply merged by scaling the NIR to the
red end of the optical flux, there could still be large discrepancies between
it and the true SED of the \sn, and one would need observed photometry to
correct for them. If the first PC were to describe some sort of correction that
adjusts the NIR prediction to be on the same true optical scale, then the
offset shown in \autoref{fig:14J_formal} is not a problem with the model, but
rather the reference observation. However, it is again difficult to confirm
this without observed photometry of the training set of \sne.

Because the differences in the flux calibration between the optical and NIR
parts of the training spectra are not completely systematic due to the
inhomogeneity of the spectroscopic observations, corrections should still be
made to the final predictions. We chose not to do so in \autoref{sec:results}
for consistency with the lack of NIR photometric observations. However, it will
yield a more accurate SED approximation with correct synthetic photometric
colors if observed photometry is available. It is also worth noting that
spectroscopic templates that are parameterized by light-curve quantities are
also photometry-corrected separately after a first approximation is given to
light-curve fitting procedures such as SNooPy.

\subsection{Comparison with Templates -- Spectroscopy}
\label{sec:templates-spectroscopy}

\begin{figure*}[ht]
    \centering
    \includegraphics[width=\textwidth]{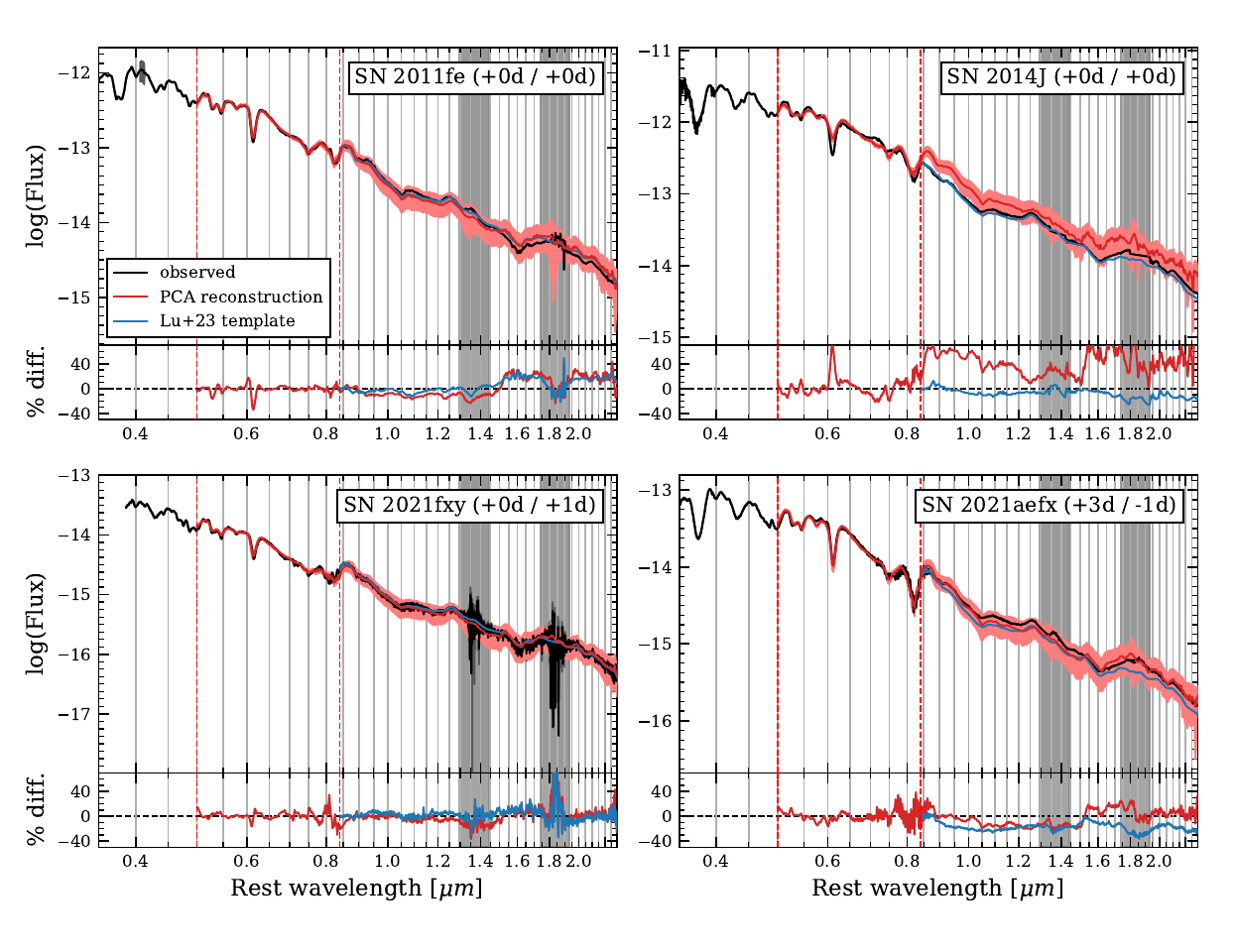}
    \caption{Comparison of near-maximum-light extrapolations of SNe 2011fe,
             2014J, 2021fxy, and 2021aefx with \citet{Lu_etal_2023} spectral
             templates given their respective values of \sbv and phase.}
    \label{fig:template_comparison}
\end{figure*}

In this section we provide a brief comparison between the extrapolation
procedure described in this work and the parameterized spectral templates
of \citet{Lu_etal_2023}.
\added{
In short, \citet{Lu_etal_2023} makes use of 331 NIR spectra of 94 \sne and
performs PCA on multiple regions in wavelength space to create a model that is
parameterized by \sbv using Gaussian process regression.
}
A substantial statistical comparison of the two methods is not provided here
due to the complexity of the predictions and lack of data with which to perform
tests.

In \autoref{fig:template_comparison}, the template approximation of SNe 2011fe,
2014J, 2021fxy, and 2021aefx are overlayed on top of the extrapolations shown
here in earlier sections. The templates were given the same phase as well
as values of \sbv for each \sn: \sbv $= 0.95$ for SN~2011fe
\citep{Ashall_etal_2019}, \sbv $= 0.99$ for SN~2021fxy
\citep{DerKacy_etal_2023fxy}, and \sbv $= 1.05$ for SN~2021aefx
\citep{Ni_etal_2023}. We also use \sbv $= 1.00$ for SN~2014J by converting its
measured value of \deltam $= 0.98$ \citep{Li_etal_2019} to \sbv using the
relation determined by \citet{Burns_etal_2018}. The template predictions were
then scaled to the optical part of the observations in the same way as the NIR
observations.

From \autoref{fig:template_comparison}, for SNe 2011fe, 2021fxy, and
2021aefx, the extrapolation and template methods produce similar results
overall. This is expected as models for both the extrapolation method and the
spectral templates were built using subsets of the same FIRE NIR spectra. In
the case of SN~2011fe, the template seems to be more true to the observation,
however after 1.5~\microns both methods deviate significantly from the
observation. As this effect is prominently seen in both the extrapolation and
the template, this may be the result of the inhomogeneity between the CSP NIR
spectra used here and the other observations, rather than the fault of the
prediction method itself. Both methods perform well for SN~2021fxy,
consistently fitting the observed spectrum in each wavelength region.
The fits for SN~2021aefx are inadequate for both methods, but
interestingly redward of 1.5~\microns the extrapolation begins to overestimate
the flux, leading to a near-zero difference in integrated flux, whereas the
template continues to underestimate the flux. The spectroscopic behavior
exhibited by SN~2021aefx and the inability of both the extrapolation and
template to reconstruct the observed spectrum may be due to
the low GMM probability for SN~2021aefx belonging to the CN Branch group. The
probability that it lies within the CN group is only 57.6\%, suggesting that it
exhibits some \SiII and other spectroscopic behavior that is atypical of
CNs.

Finally, it is clear that the template prediction is a drastically better fit
to the observation of SN~2014J than that given by the extrapolation, even if
the extrapolation were to be corrected by a constant factor. This is likely due
to the training sample used by \citet{Lu_etal_2023} being more representative
of \sne similar to SN~2014J and even BLs, as there were far fewer constraints
on data selection. It is surprising that the template produces such a good fit
to this spectrum, given that SN~2014J seems also atypical for CN \sne as it
has such a high value of \vsi. Although these four \sne all exhibit similar
values of \sbv with $0.95 <$ \sbv $< 1.05$, they each display fairly unique
spectroscopic behavior. Because of this, it is also unexpected that the
template method seems to perform better overall compared to the extrapolation
method, which makes predictions using spectroscopic information instead of the
single \sbv parameter. The likely explanation for this is the much more limited
samples sizes used for the extrapolation method, and therefore to improve this
method to be consistently on par with spectral templates, more concurrent
spectroscopic observations in both the optical and NIR are required.

\added{
Because the template method and this extrapolation method both make use of PCA
to generate models, a similar sample size of around 100 \sne (rather than the
maximum of 34 used here) would likely be sufficient in capturing the
spectroscopic information needed in both the optical and NIR to make models
that are consistently comparable to the spectral template method. It may be the
case that training these models on a comparable sample size of spectra allows
the advantages of the extrapolation method to produce better results in some
cases. Despite such a drastically smaller sample size, the extrapolation method
does well in rivaling the spectral template results.
}

\subsection{Comparison with Templates -- Photometry}
\label{sec:templates-photometry}

\begin{figure*}[ht]
    \centering
    \includegraphics[width=\textwidth]{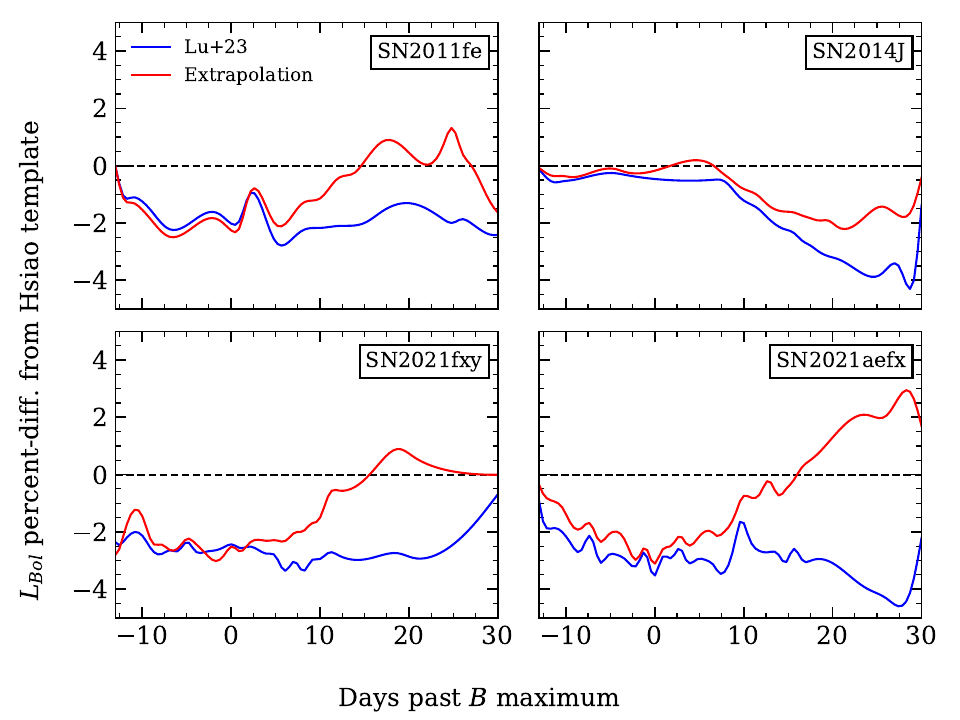}
    \caption{Percent-difference of calculated $\text{L}_{Bol}$ light-curves
             from using the \citet{Hsiao_etal_2007} template for the method
             presented in this work (Extrapolation) and for the
             \citet{Lu_etal_2023} (NIR) template. Note that the
             \citet{Hsiao_etal_2007} template does not yield light-curves
             taken as truth, but rather it is used as a reference to compare
             the two other methods.}
    \label{fig:bolometric}
\end{figure*}

Similar to \autoref{sec:templates-spectroscopy}, we also provide in this
section a comparison between the extrapolation and the spectral templates of
\citet{Lu_etal_2023} in the context of photometry. In particular, we make this
comparison by comparing bolometric light-curves obtained from SNooPy, which
has been modified slightly to reflect the procedure discussed below. This
comparison is a more realistic and applicable approach to comparing integrated
flux as a function of time.

For the same four \sne as in \autoref{sec:templates-spectroscopy} --- SNe
2011fe, 2014J, 2021fxy, and 2021aefx --- we calculate the bolometric
light-curves of each \sn in the following way. First, we gather $BVR$
photometry of SN~2011fe \citep{Richmond_Smith_2012}, $BVR$ photometry of
SN~2014J \citep{Foley_etal_2014}, $uBVgri$ photometry of SN~2021fxy
\citep{DerKacy_etal_2023fxy}, and $BVgri$ photometry of SN~2021aefx
\citep{Hosseinzadeh_etal_2022_21aefx}. This photometry allows for a
measurement of \sbv using SNooPy's \texttt{EBV\char`_model2} which in turn is
able to produce an approximate SED from the
spectral templates of \citet{Hsiao_etal_2007}, which we will henceforth refer
to as the Hsiao templates, that are built into SNooPy. From here,
three different methods are used to determine bolometric light-curves for each
\sn: the default SNooPy method, which integrates the Hsiao templates to
calculate the bolometric luminosity (\LBol); the extrapolation
method, which does the
same with the extrapolation procedure discussed in this work; and the NIR
template method, which again does the same, but using the spectral template of
\citet{Lu_etal_2023}.
\added{
For clarification, we emphasize that \LBol is calculated by integrating an
estimated SED as opposed to using the aforementioned photometric observations
of the four \sne with differing bands available. The listed photometry is
primarily used to determine \sbv, which parameterizes the spectral templates,
which are then integrated to achieve \LBol.
}

For the extrapolation and NIR template methods, the Hsiao template is used in
the optical regime until the red-most wavelength in the effective range of the
red-most band of the photometry given. For example, this is about 9000~\AA\ when
$BVR$ photometry is given. Redward of this wavelength, either the extrapolation
or NIR template method replaces the Hsiao template by merging the SEDs between
8750--9250~\AA. For the extrapolation method, the optical part of the Hsiao
template is treated as an observation and projected onto the relevant
eigenvectors so that a prediction can be made. This is necessary for testing
purposes as an observed spectrum of each supernova is not available for each
epoch that \LBol is calculated. After the full optical+NIR
SED is approximated, it follows the same process in SNooPy by correcting for
the observed photometry provided and then computing the bolometric luminosity.

The three bolometric light-curves for each \sn were calculated between $-16$ and
$+30$ days relative to $B$-maximum, and, for the purpose of comparison, the
percent-difference of \LBol given the extrapolation and NIR template methods
from that of the Hsiao templates is shown in \autoref{fig:bolometric}.
Note that the Hsiao templates and their resulting light-curves are not taken as
truth, and this difference is displayed to give them a shared point of
reference. From \autoref{fig:bolometric} we see that in \LBol space, the two
SED estimation methods yield very similar light-curves between around $-16$ to
$+10$ days until they begin to deviate consistently for each \sn. After $+10$ days
the extrapolation method yields around $+2$\% or higher bolometric luminosity
than the NIR templates when compared to the Hsiao templates. This could be due
to the lower sample sizes available at later times for the extrapolation
method (see \autoref{fig:sample_times_true}).

Finally, it is worth noting that, although the extrapolated spectrum of
SN~2014J exhibits sizable offset from the NIR template (seen in
\autoref{fig:template_comparison}), the bolometric light-curves show very small
differences between the SED estimates around maximum light. This is likely
caused by differences between the spectrum used to generate the extrapolation
shown in \autoref{fig:template_comparison} and the optical part of the Hsiao
template used to generate the extrapolations that yield light-curves from
\autoref{fig:bolometric}. In addition, the merging process occurs in different
wavelength regions. Therefore, there are no real inconsistencies in what
is seen between \autoref{fig:template_comparison} and \autoref{fig:bolometric}.
Ultimately, we find that the two methods yield very similar results in both
maximum-light spectroscopy and bolometric luminosity surrounding maximum light.
Most differences that occur between the two methods are at times later than
$+10$ days where sample sizes are smaller, leading to a less
representative sample of the population of \sne with which to train models.

\section{Conclusions}

In this study, we make use of optical spectra alongside NIR FIRE spectra from
the CSP I \& II data set to study correlations between optical and NIR
spectroscopy in an attempt to model optical-NIR behavior using PCA. This
differs from spectral template methods as this method is not intrinsically
parameterized by \sbv or other stretch-like quantities, as \sbv does not
capture the full diversity of \sne in the NIR regime. It is shown here that
using patterns exhibited by \sn spectra to extrapolate these spectra into the
NIR up to 2.3~\microns results in predictions that are in
accordance with observations. Our results show absolute differences typically
of 3\% or less in integrated flux from observations, although this may vary
greatly depending on the supernova, as is evident from the SN~2014J trial.
In cases such as this one, we find that there are large offsets between
extrapolated flux and observed flux. These offsets may be due to differences
in the flux calibrations between optical and NIR spectra that are unaccounted
for in this study. This problem may be addressed by correcting both training
and testing spectra by matching the colors of the synthetic photometry and
observed photometry at concurrent phases. This as well as large uncertainties
shown for these extrapolations clearly illustrates the need for more
observations to train these data-informed models. Specifically, the importance
of \textit{concurrent} optical and NIR spectroscopy cannot be overstated for
the study of \sne at early phases due to the variation in their spectra at
these times.

We also find that \SiII plays a significant role in partially explaining the
diversity of NIR \sn spectroscopy, which may be expected given the
well-studied
\citep{branchcomp206} classification system. Performing the extrapolation
process given only \SiII information results in similar and sometimes better
fits to observations than using a wider range of optical spectra. In addition,
projecting \SiII onto the set of maximum-light eigenvectors gives a set of
projection values that are shown to contain information that is similar to
a supernova's Branch classification, however they also contain \SiII velocity
information as well. This velocity information may be useful in explaining the
dispersion seen in the \Rline pEW-velocity relation that leads to
such high-velocity CN \sne. These projections or a similar
quantity may be good candidates to replace pEW measurements, as obtaining pEWs
can be inconsistent, since they are easily contaminated due to line blending. It
is also seen that these projections extract the SNooPy measurement of \sbv well
from a simple linear model, within a spread of about $\pm 0.1$ units.
Interestingly, this provides a method of predicting the photometric \sbv
for a \sn given only a maximum-light spectrum, therefore the maximum-light PCs
contain enough information to sub-categorize \sne. The photometric quantity
$s_{BV}$ is obtained roughly over the first 30 days from the $B$ and $V$ light
curves.
Nevertheless, a single maximum-light spectrum contains enough information to
estimate \sbv to an accuracy of $\sim$10\%, indicating that the maximum
light spectrum
contains the explosion history of the ejecta. This shows the complementarity of
spectroscopy and photometry for understanding complex astrophysical phenomena.

Finally, we compare this extrapolation method to the NIR templates of
\citet{Lu_etal_2023} and find that, although the two yield similar results
for practical purposes, the NIR templates still have an edge on extrapolation.
NIR templates give SED approximations that were slightly closer to observation
in most cases.
\replaced{
Sometimes NIR templates were much closer to observation, such as with the
particularly fast CN SN~2014J.
}{
In one case shown here, the study of SN~2014J, the NIR template was much closer
to the observation.
}
It is also seen that, when applied to bolometric light-curves, these two
methods produce consistently similar light-curves between around $-16$ to $+10$
days relative to maximum light. We speculate that, with additional data added
to the training sample, this method could indeed be capable of fitting to NIR
spectroscopic observations of \sne that contain fundamental differences that
are independent of light-curve shape. This is because we have shown that
projections onto the PCs determined here appear to have some relationship with
\sbv and spectroscopic quantities such as \SiII pseudo-equivalent widths and
\Rline velocity.
\added{
We surmise that, if future observations allow for a comparable sample size as
that used by \citet{Lu_etal_2023} ($N \sim 100$), the flexibility of this
non-parametric extrapolation method may allow for improvements compared to the
spectral template in some cases. It offers a solution to the fact that NIR
light-curves, especially around the secondary maximum, cannot be parameterized
by \sbv; with more data, this can be tested more completely.
This method is limited by the data available, but it is shown here to have
potential in predicting the NIR behavior of \sne.
}

\replaced{
This extrapolation method along with precalculated time-dependent eigenvectors
is openly maintained at \url{https://github.com/anthonyburrow/SNEx}.
}{
This extrapolation method along with precalculated time-dependent eigenvectors
is openly maintained in the software package named SNEx
\citep{Burrow_2024_SNEx}.
}
This software
takes a spectrum as input, and with user-defined parameters such as phase and
the projection region itself, the spectrum is projected onto a set of
eigenvectors to produce the resulting linear combination spectrum. There are
many variables in the work described here, many of which were assumed in this
paper for simplicity and not studied in detail. Therefore, some parameters may
yield better results when changed for certain \sne, which could be the subject
of future work when a more extensive training data sample is established.

\section{Acknowledgments}

A.B., E.B., P.B., and P.H. acknowledge support from NASA grant 80NSSC20K0538.
E.B., C.A., J.D., and  P.H. acknowledge support from NASA grants JWST-GO-02114,
\replaced{
JWST-GO-02122 and JWST-GO-04522. Support for programs \#2114, \#2122, and \#4522
}{
JWST-GO-02122, JWST-GO-04522, JWST-GO-04217, and JWST-GO-04436. Support for
programs \#2114, \#2122, \#4522 \#4217, and \#4436
}
were provided by NASA through a grant from the Space Telescope Science
Institute, which is operated by the Association of Universities for Research in
Astronomy, Inc., under NASA contract NAS 5-03127.
\added{
E.B.,  C.A., and J.D. acknowledge support from HST-AR-17555, Support for
Program number 17555 was provided through a grant from the STScI under NASA
contract NAS5-26555.
}
J.L. acknowledges support from NSF grant AAG-2206523.
L.G. acknowledges financial support from the Spanish Ministerio de
Ciencia e Innovaci\'on (MCIN), the Agencia Estatal de Investigaci\'on (AEI)
10.13039/501100011033, and the European Social Fund (ESF) "Investing in your
future" under the 2019 Ram\'on y Cajal program RYC2019-027683-I and the
PID2020-115253GA-I00 HOSTFLOWS project, from Centro Superior de Investigaciones
Cient\'ificas (CSIC) under the PIE project 20215AT016, and the program Unidad
de Excelencia Mar\'ia de Maeztu CEX2020-001058-M, and from the Departament de
Recerca i Universitats de la Generalitat de Catalunya through the
2021-SGR-01270 grant.
M.D. Stritzinger is funded by the Independent Research Fund Denmark (IRFD,
grant number 10.46540/2032-00022B).
Some of the calculations presented here were performed at the National Energy
Research Supercomputer Center (NERSC), which is supported by the Office of
Science of the U.S. Department of Energy under Contract No. DE-AC03-76SF00098,
at the H\"ochstleistungs Rechenzentrum Nord (HLRN), and at the OU
Supercomputing Center for Education \& Research (OSCER) at the University of
Oklahoma (OU). We thank all these institutions for a generous allocation of
computer time.

\software{
SNEx \citep{Burrow_2024_SNEx},
EMPCA \citep{Bailey_2012},
SNooPy \citep{Burns_CSP11},
Spextractor \citep{Burrow_etal_2020},
scikit-learn \citep[version 0.22.2,][]{scikit},
NumPy \citep[version 1.18.2,][]{numpy1,numpy2},
Matplotlib \citep[version 3.2.1,][]{matplot}
}

\clearpage

% \appendix

\bibliography{main}

% \listofchanges

\end{document}